Banner appropriate to article type will appear here in typeset article

# Turbulent flows over porous lattices: alteration of near-wall turbulence and pore-flow amplitude modulation

**Seyed Morteza Habibi Khorasani[1]†, Mitul Luhar[2] and Shervin Bagheri[1]‡**

[1]FLOW, Department of Engineering Mechanics, KTH Royal Institute of Technology, SE-100 44 Stockholm, Sweden

[2]Department of Aerospace and Mechanical Engineering, University of Southern California, Los Angeles, CA 90089, USA

Turbulent flows over porous lattices consisting of rectangular cuboid pores are investigated using scale-resolving direct numerical simulations. Beyond a certain threshold which is primarily determined by the wall-normal Darcy permeability, $K_y^+$, near-wall turbulence transitions from its canonical regime, marked by the presence of streak-like structures, to another marked by the presence of spanwise coherent structures reminiscent of the Kelvin-Helmholtz (K-H) type of instability. This permeability threshold agrees well with that previously established in studies where permeable-wall boundary conditions had been used as surrogates for a porous substrate. None of the substrates investigated demonstrate any drag reduction relative to smooth-wall turbulent flow. At the permeable surface, a significant component of the flow is that which adheres to the pore geometry and undergoes amplitude modulation (AM). This pore-coherent flow remains notable within the substrates, highlighting the importance of the porous substrate's microstructure when the overlying flow is turbulent, an aspect which cannot be accounted for when using continuum-based approaches to model porous media flows or effective representations such as wall boundary conditions. The severity of the AM is enhanced in the K-H-like regime, which has implications when designing porous substrates for transport processes. This suggests that the surface of the substrate can have a geometry which is different than the rest of it and tailored to influence the overlying flow in a particular way.

## 1. Introduction

The interaction between fluid flows and porous media has in recent years witnessed a notable increase in attention. This is owing to their wide-ranging practical applications from heat exchangers (Kuruneru, Vafai, Sauret & Gu 2020) to nuclear reactors (Hassan & Dominguez-Ontiveros 2008) as well as their pronounced presence in geophysical flows (Kazemifar, Blois, Aybar, Perez Calleja, Nerenberg, Sinha, Hardy, Best, Sambrook S. & Christensen 2021). They are also interesting from a purely flow physics perspective, particularly with regards to turbulent flows, as their permeable quality can lead to the alteration of turbulence at the interface of porous media (Manes, Poggi & Ridolfi 2011). The interactions in this interfacial flow region affect the transport processes occurring between the surface and subsurface flows, motivating the need to better understand them.

† Email address for correspondence: smhk2@mech.kth.se
‡ Email address for correspondence: shervin@mech.kth.se

Early studies of turbulence over porous media were mainly concerned with the effect that a permeable surface had on the structure of the overlying turbulent flow. This was primarily motivated by the understanding that the effect of wall permeability is distinct from wall roughness, an observation made in the comparative experiments of Zagni & Smith (1976) and Manes, Pokrajac, McEwan & Nikora (2009). One of the early numerical efforts pursuing this line of inquiry was that of Breugem, Boersma & Uittenbogaard (2006) who conducted direct numerical simulations (DNS) of turbulent channel flow with a porous substrate on one side. They classified the substrates in terms of their permeability Reynolds number, $Re_K = \sqrt{K}u_\tau/\nu = \sqrt{K}^+$, and resolved the flow inside the porous region as a continuum using the volume-averaged Navier-Stokes (VANS) equations to only recover the scales of motion which were larger than the characteristic pore dimensions. Their results demonstrated that for highly permeable cases ($\sqrt{K}^+ \gg 1$), the canonical streaks and quasi-streamwise vortices of near-wall turbulence cease to exist and are replaced by large cross-stream vortical structures. Other observations made were that the mean velocity profiles over highly permeable surfaces differed from that over impenetrable smooth walls in terms of the von Kármán constant, $\kappa$, and an also apparent lack of outer-layer similarity. Such observations were also made in the later DNS study of Kuwata & Suga (2017). The experimental work of Manes et al. (2011) however suggested that values of $\kappa$ which deviate significantly from those reported for smooth walls were due to poor inner-outer scale separation, and potentially due the low Reynolds numbers at which the aforementioned numerical studies were conducted. The lack of outer-layer similarity in those studies was also attributed to this reason. While not the focus of their work, the experiments of Kim, Blois, Best & Christensen (2020) support the conclusion made by Manes et al. (2011). The open-channel DNS of Shen, Yuan & Phanikumar (2020) which were conducted at a relatively low $Re_\tau \approx 395$ resulted in $\kappa$ values similar to those reported by Manes et al. (2011). Moreover, in the recent DNS study by Shahzad, Hickel & Modesti (2023) of turbulent flows over acoustic liners at $Re_\tau = 500 - 2000$, all flow cases demonstrated outer-layer similarity and $\kappa \approx 0.39$. Similarly to Manes et al. (2011), the latter authors attributed the $\kappa$ discrepancy in the works of Breugem et al. (2006) and Kuwata & Suga (2017) to either the low Reynolds number of those studies or the lack of pore-scale resolving simulations.

Concerning the flow inside porous substrates subject to an overlying turbulent flow, Breugem et al. (2006) observed the presence of velocity fluctuations beneath the surface and considered them to be motions induced by pressure fluctuations. Kuwata & Suga (2017) likewise attributed the sub-surface velocity fluctuations to the strong pressure diffusion directed into the substrate and caused by the K-H-like cross-stream rollers. More recently, Kim et al. (2020) provided direct experimental evidence of amplitude modulation for turbulence over permeable walls, a possibility that was highlighted by Efstathiou & Luhar (2018) based on skewness measurements in turbulent boundary layers over porous foams. This latter aspect will be of significance when examining the flow within substrates.

Kuwata & Suga (2017) examined the significance of anisotropy in porous media comprising cubic pore arrays and having either one ($K_y$), two ($K_y$, $K_z$; $K_x$, $K_y$) or three ($K_x$, $K_y$, $K_z$) non-zero diagonal components in the permeability tensor. They showed that wall-normal permeability alone did not affect the overlying turbulent flow and the additional presence of streamwise or spanwise permeability is necessary for alterations to occur. However, in the DNS study of turbulent flows over acoustic liners –perforated plates with non-zero permeability only along the wall-normal direction– by Shahzad, Hickel & Modesti (2023) it was demonstrated that such porous structures can affect the overlying flow. Further examination of anisotropic permeability was carried out in the DNS study by Gómez-de Segura & García-Mayoral (2019), where the effect of a permeable wall was incorporated using wall boundary conditions derived from the Darcy-Brinkman equation for modeling porous media flows. For highly streamwise-preferential configurations ($K_x \gg K_y$), it was shown that the wall-normal permeability is the principal component responsible for the breakdown of the near-wall streak cycle and the emergence of the spanwise coherent structures associated with the Kelvin-Helmholtz-type instability observed by Breugem *et al.* (2006). The threshold marking the onset of this K-H-like regime was estimated to be $\sqrt{K_y}^+ \approx 0.4$, with the flow falling into the fully unstable regime beyond $\sqrt{K_y}^+ \approx 0.6$. In this regime, drag performance became degraded compared to smooth-wall turbulence. Reductions in drag however where demonstrated for cases where the wall-normal permeability was $\sqrt{K_y}^+ < 0.4$.

The purpose of this study is to more closely examine the interaction between porous substrates and turbulent flows, both in the vicinity of the permeable surface and inside the porous substrate. This is investigated numerically using DNS which resolves the scales of motion from the bulk flow down to the pore-scale. The substrates examined span both the canonical wall turbulence and K-H-like regimes. This allows for determining which aspects of the flow are mainly due to the change in turbulence and whether there are flow features which persist across both flow regimes, particularly in the substrates. The effect of surface geometry is also investigated.

The structure of the paper is as follows. In §2, the porous substrate geometries considered and their characteristics along with the numerical methods are introduced. How the overlying bulk turbulence becomes modified due to the presence of substrates and the resulting consequences in terms of drag are discussed in §3, where an assessment of outer-layer similarity and characterization of the bulk turbulence regime with respect to permeability is also described. In §4, the surface flow is examined to highlight the differences that exist in terms of flow structure over the substrates falling into different turbulence regimes. The flow in the substrates is detailed in §5, and §6 discusses the amplitude modulation of this flow by the overlying turbulence. The results are summarized and discussed in §7.

## 2. DNS of turbulent flow over porous substrates

### 2.1. *Numerical method*

The configuration used in this study is an open-channel as depicted in figure 1. It is comprised of a bulk flow region with height $\delta$ and a porous substrate region of depth $h$. The direct numerical simulations using this set-up were conducted using the open-source solver PARIS simulator (Aniszewski *et al.* 2019) with in-house modifications. The code solves the incompressible Navier-Stokes equations,

$$\frac{\partial \boldsymbol{u}}{\partial t} + \boldsymbol{u} \cdot \nabla \boldsymbol{u} = -\nabla p + \frac{1}{Re_b}\nabla^2 \boldsymbol{u}, \tag{2.1a}$$

$$\nabla \cdot \boldsymbol{u} = 0, \tag{2.1b}$$

where $\boldsymbol{u} = (u, v, w)$ and $p$ are the velocity vector and pressure, respectively. The bulk Reynolds number is $Re_b = U_b\delta/\nu$ with $U_b$ being the bulk mean velocity in $0 < y < \delta$, and $\nu$ the kinematic viscosity.

The dimensions of the computational domain are $(L_x, L_y, L_z) = (6.3\delta, 1.3\delta, 3.15\delta)$ in the streamwise ($x$), wall-normal ($y$) and spanwise ($z$) directions. The boundaries are periodic in $x$ and $z$ with symmetry and no-slip boundaries imposed at the upper and lower domain boundaries along $y$, respectively. The thickness of the porous substrate is $h = 0.3\delta$. The simulations were conducted at a constant mass flow rate, which was adjusted to achieve a nominal $Re_\tau = u_\tau \delta / \nu = 360$ for most of the cases. Here, $u_\tau = \sqrt{\tau_w / \rho}$ is the friction velocity at $y = 0$ (the substrate surface). Note that the total stress at the surface plane, $\tau_w$, will have both viscous and Reynolds shear stress components due to the surface permeability, therefore

$$\tau_w = \mu \frac{\partial u}{\partial y}\bigg|_{y=0} + \rho \overline{u'v'}|_{y=0}. \tag{2.2}$$

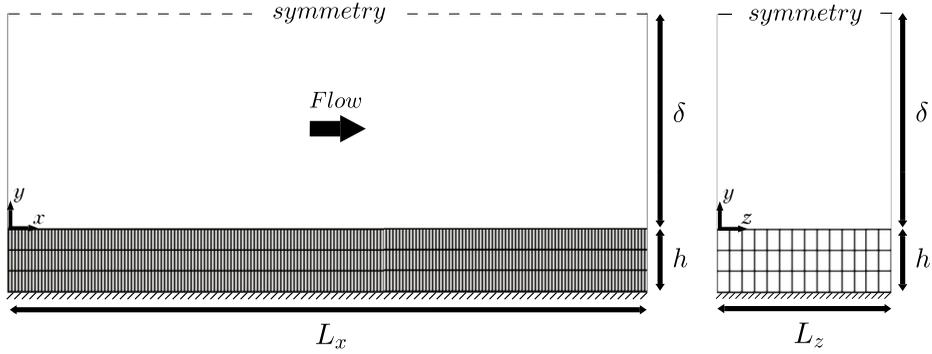

Figure 1: Sketch of the computational domain.

Throughout this paper, the superscript '+' indicates inner units, where quantities are normalized using the friction velocity and kinematic viscosity. The numerical grid has a resolution of $(N_x, N_y, N_z) = (1620, 324, 810)$. The grid spacing is uniform along $x$ and $z$ while being non-uniform along $y$. The grid is stretched in the region above the substrate using a hyperbolic tangent function and uniform within the substrate. The simulation results are grid independent at the chosen resolution. The grid independence was determined by conducting simulations at coarser and finer resolutions. The results of these simulations are gathered in appendix A.

The equations are spatially discretized on a staggered Cartesian grid using central second-order finite-differencing. The fractional-step method (Kim & Moin 1985) is used to solve the discretized incompressible Navier-Stokes equations. At each time step an intermediate non-divergence free velocity field is first calculated. The Poisson equation obtained by imposing the incompressibility constraint is then solved in Fourier space using a FFT-based solver (Costa 2018) to obtain the pressure correction. The pressure correction is then used to project the velocity field onto a divergence free vector space. The time integration uses a triple-substep Runge-Kutta method where both the advective and diffusive terms are treated explicitly.

The immersed boundary method of Breugem & Boersma (2005) is used to numerically realize the porous substrates. Unlike direct-forcing or penalization-based methods, it involves modifying the discretized advective and diffusive flux terms of the Navier-Stokes equations such that the no-slip and no-penetration conditions become exactly imposed for the regions of the numerical domain which are defined as solids. This makes it a more accurate method for realizing geometries which are Cartesian conforming, as demonstrated by Paravento, Pourquie & Boersma (2008). This IBM along with PARIS have also been used in the DNS study of turbulent flows over liquid-infused surfaces of Sundin, Zaleski & Bagheri (2021).

In addition to the simulations including porous substrates, an open-channel flow over a smooth wall at $Re_\tau = 360$ was also simulated to serve as a baseline for comparisons.

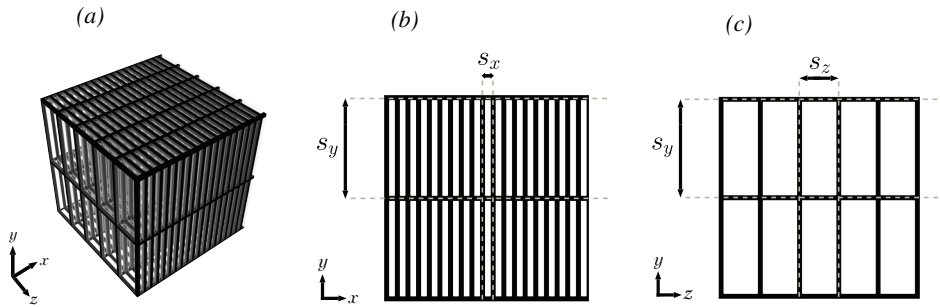

Figure 2: *(a)* Schematic of general substrate geometry along with *(b, c)* its nomenclature.

Quantities normalized using the smooth-wall friction velocity and kinematic viscosity are indicated by the subscript '*o*'.

### 2.2. *Porous geometries*

The porous geometries investigated are lattices where the pores are repeating rectangular cuboids (figure 2). The solid skeleton is made of rods with a cross-section of $d \times d$. The spacings (pitch-lengths) $s_x$, $s_y$ and $s_z$ determine the cross-sections of the pores and their resulting void volume. The components of the resulting anisotropic permeability tensor, $\boldsymbol{K}$, were calculated by conducting Stokes flow simulations of representative volumes of each substrate with the same solver that was used for the turbulent flow simulations. The dimensions of the REVs were $(L_x, L_y, L_z) = (0.3\delta, 0.3\delta, 0.3\delta)$ and the grid resolution similar to the turbulent simulations in terms of spacing. The components of the permeability tensor were then obtained by using Darcy's law,

$$\langle \boldsymbol{u} \rangle = -\frac{1}{\mu} \boldsymbol{K} \, \nabla \langle p \rangle^f. \tag{2.3}$$

As similarly defined by Breugem *et al.* (2006), $\langle \boldsymbol{u} \rangle$ and $\nabla \langle p \rangle^f$ are the superficially volume-averaged velocity and intrinsically (fluid-phase) volume-averaged pressure, respectively. Intrinsic averages are linearly related to their superficial counterparts, such that $\langle \cdot \rangle^f = \langle \cdot \rangle / \varphi$. Here, the porosity, $\varphi$, represents the void fraction of a substrate's volume.

Similar to porous structures studied by Kuwata & Suga (2017), the permeability tensor becomes diagonal for the geometries under consideration here due to the symmetry along the $x$, $y$ and $z$ directions, giving the three permeability components $K_x$, $K_y$ and $K_z$. The lengths $\sqrt{K_x}$, $\sqrt{K_y}$ and $\sqrt{K_z}$ then characterize each porous geometry. This choice of characterization is made to remain consistent with that of prior experimental and numerical studies (Manes *et al.* 2011; Kuwata & Suga 2017; Gómez-de Segura & García-Mayoral 2019). Additionally, since the Darcy permeabilities are defined in the Stokes regime, they are only dependent on the geometry and are therefore suitable parameters for encapsulating the geometrical features of different porous media. The main cases ($HP1$, $HP2$, $HP3$, $MP$, $LP1$, $LP2$, $LP3$) span a range of wall-normal permeabilities, $K_y$, to facilitate investigating how near-wall turbulence becomes altered due to a progressively weakening wall-impedance. $HP$ designates *Higher Permeability* substrates ($\sqrt{K_y}^+ > 2$), $LP$ designates *Lower Permeability* substrates ($\sqrt{K_y}^+ < 1$), and $MP$ a *moderate permeability* case residing between the other two groups $1 < \sqrt{K_y}^+ < 2$. Cases $HP2'$ and $HP3'$ –where the wall-parallel spacings of $HP2$ and $HP3$ have been swapped– retain the wall-normal permeability of $HP2$ and $HP3$ but have increased streamwise permeability and hence anisotropy, $\Phi_{xy} = K_x/K_y$.

| Case | Symbol | $Re_\tau$ | $\varphi$ | $s_{x_o}^+$ | $s_{y_o}^+$ | $s_{z_o}^+$ | $\sqrt{K_{x_o}^+}$ | $\sqrt{K_{y_o}^+}$ | $\sqrt{K_{z_o}^+}$ | $\Phi_{xy}$ |
|------|--------|-----------|-----------|-------------|-------------|-------------|--------------------|--------------------|--------------------|-------------|
| *HP1* | ◯ | 385 | 0.75 | 36.0 | 108.0 | 28.0 | 3.21 | 3.44 | 4.72 | 0.93 |
| *HP2* | ☆ | 373 | 0.68 | 36.0 | 54.0 | 28.0 | 2.74 | 2.62 | 3.84 | 1.05 |
| *HP3* | ◇ | 360 | 0.62 | 36.0 | 36.0 | 28.0 | 2.19 | 2.19 | 2.81 | 1.00 |
| *MP* | ◯ | 359 | 0.50 | 28.0 | 27.0 | 28.0 | 1.50 | 1.53 | 1.50 | 0.98 |
| *LP1* | ☆ | 356 | 0.49 | 21.0 | 54.0 | 21.0 | 1.04 | 0.73 | 1.04 | 1.43 |
| *LP2* | ◇ | 359 | 0.39 | 21.0 | 36.0 | 21.0 | 0.91 | 0.62 | 0.91 | 1.47 |
| *LP3* | ▢ | 359 | 0.26 | 21.0 | 21.0 | 21.0 | 0.50 | 0.50 | 0.50 | 1.00 |
| | | | | | | | | | | |
| *HP2′* | ☆ | 378 | 0.68 | 28.0 | 54.0 | 36.0 | 3.84 | 2.62 | 2.74 | 1.47 |
| *HP3′* | ◇ | 361 | 0.62 | 28.0 | 36.0 | 36.0 | 2.81 | 2.19 | 2.19 | 1.28 |

Table 1: DNS cases of open-channel turbulent flow over porous substrates. The porosity for each substrate is given by $\varphi$. The pore spacings are $s_{x_o}^+$, $s_{y_o}^+$ and $s_{z_o}^+$, while $\sqrt{K_{x_o}^+}$, $\sqrt{K_{y_o}^+}$ and $\sqrt{K_{z_o}^+}$ are the effective permeabilites which are analogous to the permeability Reynolds number, $Re_K$. The ratio of streamwise to wall-normal permeability is $\Phi_{xy}$. The rod or filament thickness of the solid matrix is $d/\delta = 0.039$ or $d_o^+ = 14$ for all cases. Labels, colors and symbols remain consistent throughout the manuscript.

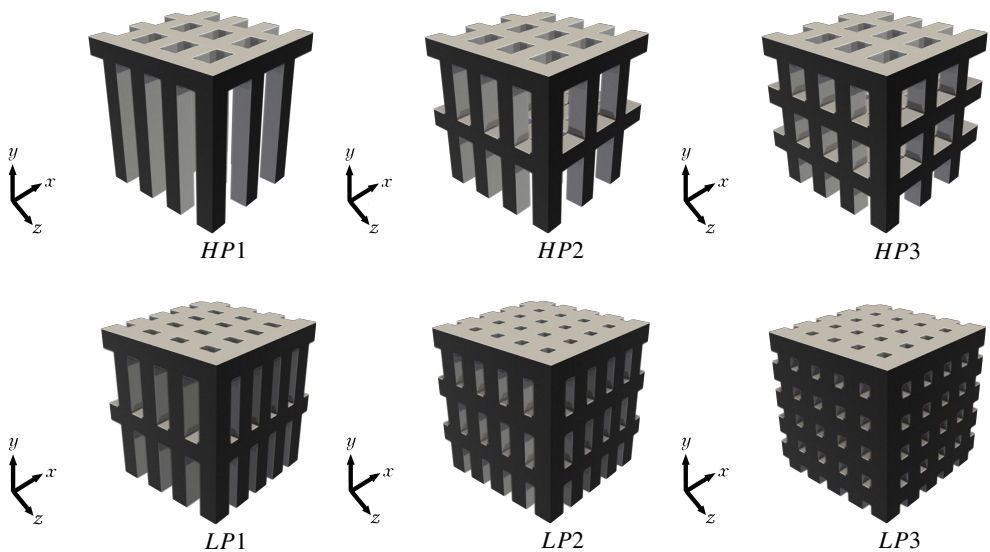

Figure 3: Representative elements of the *LP* and *HP* substrates in table 1. *HP2′* and *HP3′* (not shown) are rotated versions of *HP2* and *HP3* around the *y* axis by $\pi/2$.

The simulation parameters along with the characteristics of their porous substrates are gathered in table 1. Representative elements for the *HP* and *LP* porous geometries examined are shown in figure 3. The substrates will first be assessed with regards to certain aspects of permeable-wall turbulence which have been reported in the literature. This includes the transition of turbulence from the canonical near-wall regime to the K-H-like regime (Gómez-de Segura & García-Mayoral 2019) and how streamwise-preferential anisotropy affects the drag (Rosti *et al.* 2018; Gómez-de Segura & García-Mayoral 2019). An important difference exists between pore-scale resolving simulations, and other simulation approaches such as volume-averaged Navier-Stokes (VANS) simulations (Breugem *et al.* 2006; Rosti *et al.*

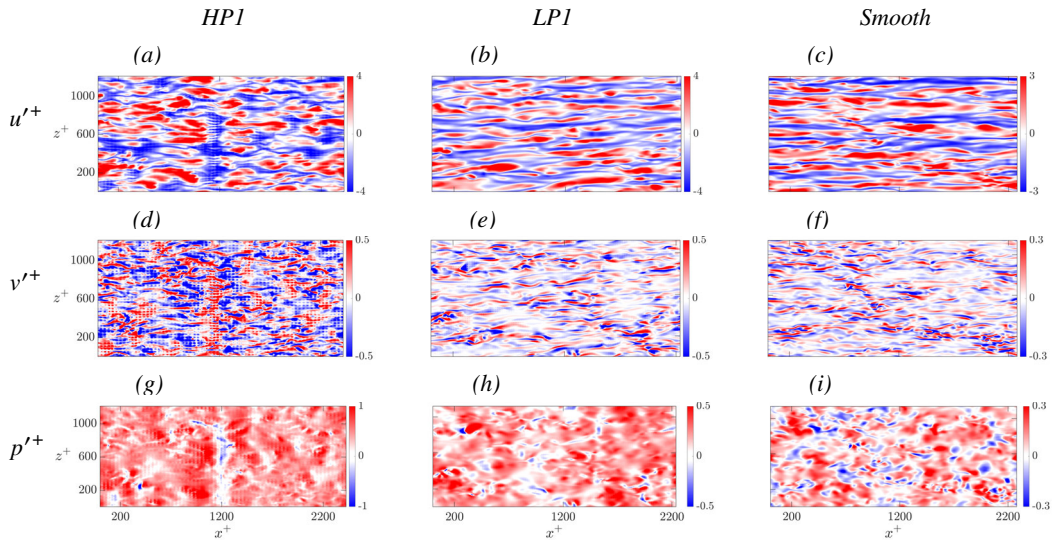

Figure 4: Instantaneous *(a, b, c)* streamwise velocity fluctuations, *(d, e, f)* wall-normal velocity fluctuations, and *(g, h, i)* pressure fluctuations at $y^+ \approx 5$. First column, $HP1$; second column, $LP1$; third column, smooth-wall. Flow direction is from left to right.

2018) and permeable-wall boundary conditions (Jiménez *et al.* 2001; Gómez-de Segura & García-Mayoral 2019). In the latter approaches, the permeability and porosity are predefined numerical parameters and independent from one another. For the cases considered in this work, the substrate's geometry determines both the permeability and the porosity. This imposes constraints on the permeabilities and the degree of anisotropy that can be obtained. For example, a change in $s_x$ will change the pore cross-section along both $y$ and $z$, leading to changes in $K_y$ and $K_z$. The Reynolds numbers of DNS studies are also generally low relative to experiments. As such, achieving effective permeabilities $\left(\sqrt{K_x}^+, \sqrt{K_y}^+, \sqrt{K_z}^+\right)$ which are large becomes challenging.

## 3. Overlying turbulent flow

### 3.1. *Surface region flow*

The changes in the overlying turbulent flow are first assessed qualitatively by examining the flow-field above the permeable surface at $y^+ \approx 5$. Figure 4 shows instantaneous snapshots of the velocity and pressure fluctuations for $HP1$ and $LP1$. These two cases have the highest $\sqrt{K_y}^+$ of the $HP$ and $LP$ groups, respectively. This allows for a better assessment of the role wall impedance plays in causing changes in turbulence near the surface. Note that at the surface, the relevant permeability component is $K_y$ as the flow must first be able to penetrate into the substrate before becoming redirected into the horizontal directions, for which $K_x$ and $K_z$ are important.

Beginning with observations for the streamwise velocity component, $LP1$ (figure 4b) exhibits regions of elongated positive and negative fluctuations which are similar to the streaky pattern observed over a smooth-wall (figure 4c). This streamwise coherency is diminished in $HP1$ (figure 4a), where the aforementioned regions are instead more clump-like and also exhibit a degree of regularity along the spanwise dimension of the domain. This is indicative of the near-wall turbulence becoming altered. For the wall-normal velocity component, the similarity between $LP1$ (figure 4e) and the smooth-wall (figure 4f) is reflective of the similarity seen in their streamwise velocity fields. $LP1$ has a greater degree of intensity but remains structurally similar to its smooth-wall counterpart. $HP1$ (figure 4d) however shows not only a noticeably stronger intensity in wall-normal velocity fluctuations but also structural differences, with there seemingly being an emergent spanwise coherency. The differences in the pressure fields are consistent with the observations made for the velocity fields. $HP1$ (figure 4g) demonstrates a spanwise patch along $x^+ \approx 1200$ which is coincident with the spanwise coherent regions seen at the same position in its $u$ and $v$ fields. The pressure fluctuations of $HP1$ are also more intense compared to $LP1$ (figure 4h) and the smooth-wall case (figure 4i). Another distinctive quality of $HP1$ is the visible signature of the permeable surface in its flow-field, particularly when examining the wall-normal velocity (figure 4d). This indicates that the surface granularity becomes perceived by the turbulent flow, such that it leaves a visible footprint in the flow-field. This is similar to what has been observed in flows over roughness (Abderrahaman-Elena *et al.* 2019) and canopies (Sharma & García-Mayoral 2020), albeit at lower heights, and is attributed to the flow induced by the surface elements. That this footprint remains visible at $y^+ \approx 5$, suggests that the surface-induced flow above $HP1$ is of notable strength.

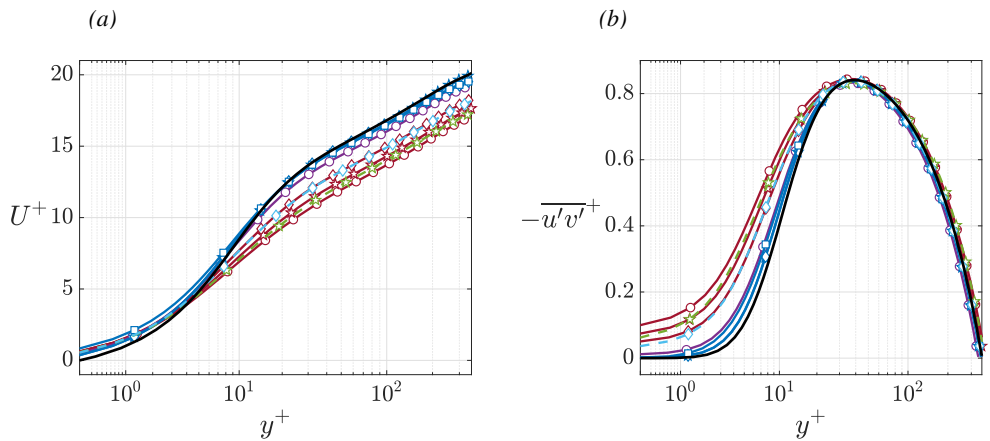

Figure 5: (*a*) Mean velocity and (*b*) Reynolds shear-stress distributions of the bulk flow overlying the porous substrates. Symbols and colors follow the descriptions in table 1. The black line is a reference smooth-wall solution at $Re_\tau = 360$.

### 3.2. *Changes in mean flow, velocity fluctuations and Reynolds shear stress*

Surfaces which depart from a hydrodynamically smooth behavior change the overall level of momentum carried by the bulk flow, i.e., the amount of drag generated at the surface changes. As explained by Spalart & McLean (2011) and Chung *et al.* (2021), a suitable metric for quantifying this is the shift in the logarithmic region of the mean velocity profile, $\Delta U^+$, which was introduced by Hama (1954) and Clauser (1954) and is more commonly known as the roughness function. For the substrates considered, this is examined in figure 5a where the mean velocity profiles are shown. Table 2 lists the corresponding $\Delta U^+$ for each configuration. All of the porous substrates examined increase drag compared to the baseline smooth-wall case, although the $LP$ cases do not impose a significant drag penalty. The structural differences observed in the flow over the different substrates are also reflected here, such that they become distinguishable into two overall groups. The substrates with lower wall-normal permeability ($LP1$, $LP2$, $LP3$, $MP$) do not deviate greatly from the smooth-wall ($-\Delta U^+ \lessapprox 1$) and the $LP$ cases are almost indistinguishable from the smooth-wall case.

The substrates with higher wall-normal permeability ($HP1$, $HP2$, $HP2'$, $HP3$, $HP3'$) on the other hand result in $\Delta U^+$ values which are notable ($-\Delta U^+ \gtrsim 2$). The same distinction can be made for the Reynolds shear stress (figure 5b), where substrates $LP1$, $LP2$, $LP3$ and $MP$ result in slightly greater levels of turbulent activity close to the permeable surface whereas this activity is more pronounced for the $HP$ substrates with a gap emerging between the former and the latter substrates in terms of their surface-level $-\overline{u'v'}^+$ activity.

Following the approach of MacDonald et al. (2016) and García-Mayoral et al. (2019), to assess the contributing factors to $\Delta U^+$, the mean momentum equation for the bulk flow region above the substrate is considered

$$\nu \frac{dU}{dy} - \overline{u'v'} = u_\tau^2 \left(1 - \frac{y}{\delta}\right). \tag{3.1}$$

Scaling (3.1) in inner units gives

$$\frac{dU^+}{dy^+} - \overline{u'v'}^+ = 1 - \frac{y^+}{\delta^+}. \tag{3.2}$$

Integrating (3.2) between $y^+ = 0$ (the substrate surface) and a position within the logarithmic region, $y^+ = H^+$, results in

$$\int_{y^+=0}^{y^+=H^+} -\overline{u'v'}^+ \, dy^+ + U^+\left(y^+ = H^+\right) - U^+\left(y^+ = 0\right) = H^+ - \frac{H^{+2}}{2\delta^+}. \tag{3.3}$$

For a smooth-wall flow, there will be no mean velocity at $y^+ = 0$ (no-slip) and hence $U^+(y^+ = 0) = U^+_{\text{slip}} = 0$. Taking the difference of (3.3) between a porous case and the smooth-wall case allows for quantifying the contributions of the interfacial slip velocity and changes in Reynolds stress to $\Delta U^+$,

$$\Delta U^+ = U^+\left(y^+ = H^+\right)_{\text{porous}} - U^+\left(y^+ = H^+\right)_{\text{smooth}}$$

$$= U^+{}_{\text{slip}} - \int_{y^+=0}^{y^+=H^+} \left[\left(-\overline{u'v'}^+_{\text{porous}}\right) - \left(-\overline{u'v'}^+_{\text{smooth}}\right)\right] dy^+$$

$$-\frac{H^{+2}}{2}\left(\frac{1}{\delta^+{}_{\text{porous}}} - \frac{1}{\delta^+{}_{\text{smooth}}}\right)$$

$$= \tau_{\text{slip}} + \tau_{uv} + \tau_{Re} \approx \tau_{\text{slip}} + \tau_{uv}. \tag{3.4}$$

The term $\tau_{Re}$ quantifies the contributions from additional turbulence scales and emerges due to the differences in $Re_\tau$, i.e. $\delta^+{}_{\text{porous}}$ and $\delta^+{}_{\text{smooth}}$. However, it remained negligible for the simulations conducted here and is therefore omitted.

In figure 6, it can be observed that the slip velocity contribution, $\tau_{\text{slip}}$, to $\Delta U^+$ does not vary significantly across the different cases. The Reynolds shear stress contribution, $\tau_{uv}$, is drag degrading and hence always negative. The magnitude of $\tau_{uv}$ decreases monotonically from $HP1$, which has the overall highest permeability, to the $LP$ cases which are similar in terms of $\Delta U^+$. The dominant component across all cases is $\tau_{uv}$ and grows larger for substrates with greater wall-normal permeabilities. It is notably larger in magnitude than $\tau_{\text{slip}}$ for the $HP$ cases. These results are consistent with what was observed for the profiles of the mean velocity (figure 5a) and Reynolds shear stress (figure 5b). A jump in $\tau_{uv}$ is also seen here when going from the $LP$ cases to the $HP$ cases, suggestive of additional contributions resulting from the structural changes in turbulence that were observed in figure 4. The same trend holds for $HP2'$ and $HP3'$ which have streamwise-preferential anisotropy unlike their baseline counterparts $HP2$ and $HP3$. However, their increased anisotropy leads to increased drag. In the *LP* cases on the other hand, as the anisotropy increases from *LP*3 to *LP*1 the drag becomes reduced, although the changes are small. Overall, none of the substrates allow for the development of any significant slip velocity at their surfaces. Therefore, the differences in $\Delta U^+$ are primarily due to the differences in $-\overline{u'v'}^+$ close to the permeable surface.

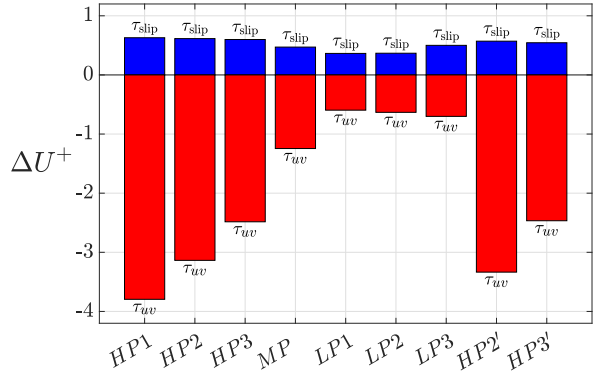

Figure 6: The slip velocity, $\tau_{\mathrm{slip}}$, and Reynolds shear stress ,$\tau_{uv}$, contributions to $\Delta U^+$.

### 3.3. *Outer layer similarity*

Throughout the literature, assessments have been made about the applicability of the logarithmic law of the wall to permeable wall turbulence with inconsistent results. The logarithmic law of the wall is

$$U^+ = \frac{1}{\kappa}\ln\left(y^+\right) + B + \Delta U^+,\qquad(3.5)$$

as introduced by Clauser (1954) for rough wall turbulence and takes into account $\Delta U^+$. The argument being that the only difference between a smooth and non-smooth turbulent flow is $\Delta U^+$ while the von Kármán constant, $\kappa$, remains the same. Commonly, when assessing the log-law over rough or permeable walls, a displacement height, $d_0$, is introduced and the wall-coordinates are changed such that $y = -d_0$ becomes the effective wall origin. Here, the approach of Orlandi & Leonardi (2006) is employed instead, where the mean velocity is measured relative to the slip velocity which exists at the permeable surface of the porous substrates, giving

$$U^+ - U_{\mathrm{slip}}^+ = \frac{1}{\kappa}\ln\left(y^+\right) + B + \left(\Delta U^+ - U_{\mathrm{slip}}^+\right).\qquad(3.6)$$

In this manner, introduction of a displacement height is avoided and the porous cases share a common definition with the smooth-wall case at the surface since the relative mean velocity there becomes zero for all of them. The values of $\kappa$ estimated by fitting the modified law of the wall (3.6) for the various cases are gathered in table 2. The estimated values fall within the range of those typically reported for smooth-wall turbulent boundary layers, and $\kappa$ is approximately the same for all examined cases. This suggests that outer-layer similarity is retained for the cases investigated here. Examining the velocity fluctuations and Reynolds shear stress distributions in outer-scaled wall-coordinates (figure 7) reinforces the existence of outer-layer similarity for the turbulent flow over all the porous substrates considered. This is in line with the DNS study of turbulent flows over acoustic liners done by Shahzad *et al.* (2023) who obtained a constant $\kappa$ value for both their acoustic liner and smooth-wall cases.

| Case | $\kappa$ | $\Delta U^+$ | B | RMSE | $R^2$ |
|------|------|------|------|------|------|
| Smooth-wall | 0.39 | 0.00 | 4.88 | 0.04 | 0.99 |
| $HP1$ | 0.38 | -3.17 | 5.31 | 0.05 | 0.99 |
| $HP2$ | 0.38 | -2.52 | 5.35 | 0.05 | 0.99 |
| $HP3$ | 0.39 | -1.89 | 5.52 | 0.05 | 0.99 |
| $MP$ | 0.39 | -0.77 | 5.53 | 0.05 | 0.99 |
| $LP1$ | 0.39 | -0.23 | 4.99 | 0.07 | 0.99 |
| $LP2$ | 0.39 | -0.27 | 5.56 | 0.07 | 0.99 |
| $LP3$ | 0.39 | -0.34 | 5.94 | 0.06 | 0.99 |
| $HP2'$ | 0.39 | -2.76 | 5.71 | 0.02 | 0.99 |
| $HP3'$ | 0.39 | -1.92 | 5.53 | 0.05 | 0.99 |

Table 2: The von Kármán constant, $\kappa$, and log layer intercept, B, resulting from the fitting of the law of the wall (3.6) to the mean velocity profiles of the different cases in table 1 along with their respective values of $\Delta U^+$. The last two columns report the root-mean-square errors and coefficient of determinations, respectively.

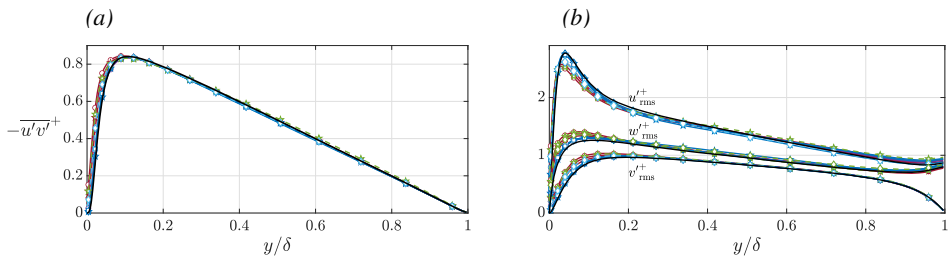

Figure 7: *(a)* Reynolds shear stress and *(b)* root mean square velocity fluctuations above the substrates in outer-scaled wall coordinates.

Prior observations made of $\kappa$ differing from its typical smooth-wall value for turbulence over porous media could therefore be due to the absence of sufficient scale-separation, such that the overlying flow is altered up to a significant distance away from the substrate region. This was also suggested by Shahzad *et al.* (2023) as the probable reason behind the discrepancy in $\kappa$ values reported by Breugem *et al.* (2006) and Kuwata & Suga (2017). Manes *et al.* (2011) and Chen & García-Mayoral (2023) suggest that the fitting approach adopted by Breugem *et al.* (2006) and Kuwata & Suga (2017) for estimating $\kappa$ could be a source of error.

### 3.4. *Flow regime distinction with regards to permeability*

The separation of the substrates into two groups becomes more clearly distinguishable when plotting the drag change, $\Delta U^+$, against the wall-normal permeability (figure 8a). The leap in $\Delta U^+$ when going from the $LP$ to the $HP$ cases has been attributed to the near-wall turbulence dynamics undergoing a transition and departing the canonical regime for a K-H-like one (Gómez-de Segura & García-Mayoral 2019). Using linear stability analysis with boundary conditions derived from the Darcy-Brinkman equation, Gómez-de Segura *et al.* (2018) proposed the following relation for quantifying the influence of a substrate's permeability on triggering this transition:

$$K_{Br_1}^+ = K_y^+ \tanh\left(\frac{\sqrt{2K_x^+}}{9}\right)\tanh^2\left(\frac{h^+}{\sqrt{12K_y^+}}\right) \approx K_y^+ \tanh\left(\frac{\sqrt{2K_x^+}}{9}\right). \tag{3.7}$$

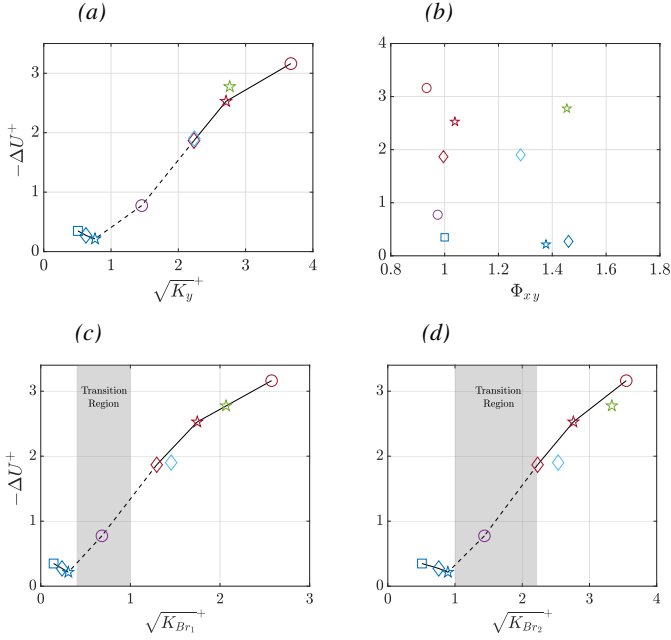

Figure 8: Mean velocity shift plotted against (*a*) wall-normal permeability, (*b*) the streamwise to wall-normal anisotropy ratio, (*c*) effective permeability of (3.7), and (*d*) effective permeability of (3.8). □, *LP*1; ◇, *LP*2; ☆, *LP*3; ○, *MP*; ◇, *HP*3; ☆, *HP*2; ○, *HP*2; ◇, *HP*3′; ☆, *HP*2′.

For sufficiently deep substrates, the second hyperbolic tangent term becomes $\approx 1$ and the relation becomes simplified, with the dominant term becoming $K_y^+$. Gómez-de Segura & García-Mayoral (2019) determined $\sqrt{K_{Br_1}}^+ \approx 0.4 - 0.6$ as the threshold in which the onset and transition to the K-H-like regime occurs. This agrees with the simulation data in this work, shown in figure 8c, where $\sqrt{K_{Br_1}}^+ \approx 0.31$ for *LP*1, and beyond which the K-H-like instability becomes triggered and subsequently grows in strength (shaded region in figure 8c). Cases *HP*2′ and *HP*3′ which differ in terms of their $\Phi_{xy}$ anisotropy from the other *HP* cases (figure 8b) also conform to (3.7). While (3.7) does seem suitably applicable to the porous substrates considered here, it was originally obtained for conditions where $\sqrt{K_x}^+ > \sqrt{K_y}^+$. In related work by Sharma *et al.* (2017), it was demonstrated that for $\sqrt{K_x}^+ < \sqrt{K_y}^+$, linear stability analysis leads to different results, expressed using

$$K_{Br_2}^+ = \sqrt{K_x^+ K_y^+} \tanh\left(\frac{h^+}{18}\sqrt{\frac{K_x^+}{K_y^+}}\right)\tanh^2\left(\frac{h^+}{\sqrt{K_x^+}}\right) \approx \sqrt{K_x^+ K_y^+}, \qquad (3.8)$$

where the hyperbolic tangent terms again become $\approx 1$. The results from applying (3.8) to the simulation data of the cases in table 1 are shown in figure 8d. They are similar to those in figure 8c obtained using (3.7). Anisotropy becomes reflected more prominently when characterising the substrates using (3.8), as *HP*2′ and *HP*3′ become more separated from *HP*2 and *HP*3 in figure 8d compared to figure 8c. Also, the change when going from *LP*3 to *MP* is captured with more abruptness when using (3.8). Sharma *et al.* (2017) reported that for $\sqrt{K_{Br_2}}^+ > 2.2$ the instability becomes fully developed. For *HP*3, $\sqrt{K_{Br_2}}^+ \approx 2.2$, and in section §3.5 it will be shown through spectral analysis that the *HP* cases belong to the K-H-like regime. As such, the instability criteria of Gómez-de Segura *et al.* (2018) and Sharma *et al.* (2017) are good predictors of when the turbulence dynamics over porous structures changes from its canonical nature. The first-order influence of $K_y{}^+$ becomes evident when taking into account that the weakening of the wall-blocking effect at the surface is directly tied to this permeability component. Once permeability at the surface is present to a sufficiently large degree to permit the penetration of momentum into the substrate, the fluid moving below the surface must then contend with the horizontal blockage imposed by the substrate which is characterized by $K_x{}^+$ and $K_z{}^+$, giving these permeability components second-order significance. The two criteria of (3.7) and (3.8) will probably exhibit more distinguishable results from one another for substrates with higher anisotropy, whereas here they are mostly similar to one another.

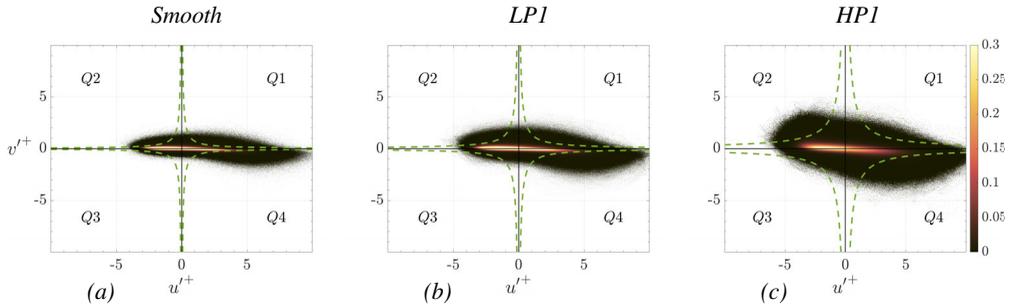

Figure 9: Quadrant maps of $u'$ and $v'$ at $y^+ \approx 5$: ($a$) smooth-wall, ($b$) $LP1$, ($c$) $HP1$. The dashed lines are hyperbolas marking $|u'^+v'^+| = 8 \times -\overline{u'v'}^+$. Color-bar indicates the frequency of events in each bin of the map.

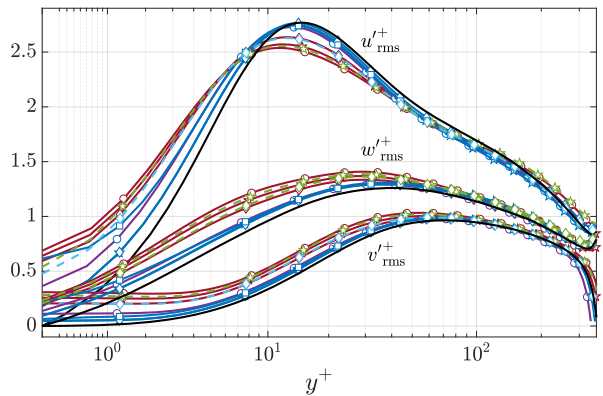

Figure 10: Distributions of the root mean square velocity fluctuations for the bulk turbulent flow overlying the porous substrates.

### 3.5. *Turbulence structure*

In figure 5b, it was shown that the $HP$ cases are distinguished by higher Reynolds shear stresses in the region close to the permeable surface. Quadrant analysis of the velocity fluctuations can be leveraged to examine the change in turbulence intensities with respect to flow events, in particular the contribution from ejections ($Q2$, $u' < 0$ and $v' > 0$) and sweeps ($Q4$, $u' > 0$ and $v' < 0$). This is shown for the $y^+ \approx 5$ plane above the surface in figure 9. Going from an impenetrable smooth-wall (figure 9a) to permeable case $LP1$ (figure 9b) and finally the greater permeable case of $HP1$ (figure 9c), a tilting and expansion of the joint probability distributions of $u'$ and $v'$ is seen. As explained by Manes *et al.* (2011), the tilting is attributable to an increase in $v'$ activity, which is also evident when viewing the r.m.s. velocity fluctuations in figure 10, particularly when going from the $LP$ to the $HP$ cases. In terms of flow events, sweeps become increasingly dominant as the permeability increases, in both strength and number of occurrences, contributing to a greater generation of Reynolds shear stress in the near-surface region. This is a feature common to flows over permeable surfaces, be they canopies (Finnigan, Shaw & Patton 2009) or porous media (Manes *et al.* 2011).

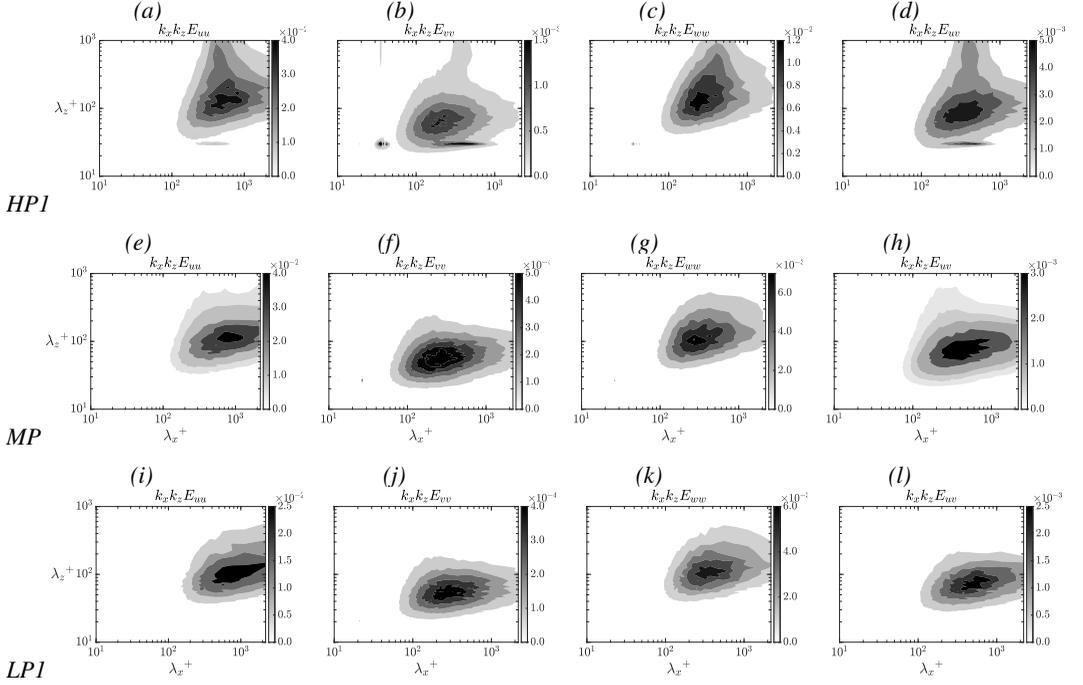

Figure 11: Pre-multiplied two-dimensional spectral energy densities: $(a, e, i)$ $k_x k_z E_{uu}$, $(b, f, j)$ $k_x k_z E_{vv}$, $(c, g, k)$ $k_x k_z E_{ww}$, and $(d, h, l)$ $k_x k_z E_{uv}$ at $y^+ \approx 5$. First row, $HP1$; second row, $MP$; third row, $LP1$.

The experiments of Manes *et al.* (2011) also showed the distribution of points initially growing and then subsequently shrinking when going from their lowest permeability to their highest permeability case (refer to figure 15 of their manuscript). They attributed this behaviour to the near-surface flow mechanism changing to a mixing-layer type for the highest permeability porous media they investigated, similar to what occurs for turbulence over canopies (Finnigan 2000). The highest permeability examined by Manes *et al.* (2011), for which they observed mixing-layer behavior, was $\sqrt{K}^+ \approx 17$ at $Re_\tau \approx 3848$. This is considerably greater than the highest effective wall-normal permeability investigated here ($\sqrt{K_y}^+ \approx 3.4$), suggesting that the cases in this study do not fall into the category of mixing-layer type behavior.

The differences in turbulence structure become better established by examining the spectral energy densities of the velocity fluctuations. The spectra of cases $HP1$ and $LP1$ in figure 11 are different form one another, with $HP1$ exhibiting energetic scales at large spanwise wavelengths which are absent in $LP1$. For case $MP$, the onset of the instability can be inferred from the emergence of energetic scales at large $\lambda_z^+$, but it is not yet intensified. This spanwise coherent component is attributable to the existence of spanwise rollers associated with a K-H-like instability over permeable boundaries (Gómez-de Segura & García-Mayoral 2019). The presence of these roller structures for the $HP$ cases can be observed in the vortex visualization of figure 12. Additionally, by using spectral proper orthogonal decomposition (SPOD), modes are obtained for $HP1$ which capture the spanwise coherent rollers but for $LP1$, and indeed all of the $LP$ cases, no such modes are obtained. These SPOD modes for $HP1$ may be viewed in appendix B, but have been omitted from the main text for brevity. The emergence of these K-H-like structures is the cause behind the intensification of turbulent activity in the proximity of the substrate surface and drag increase.

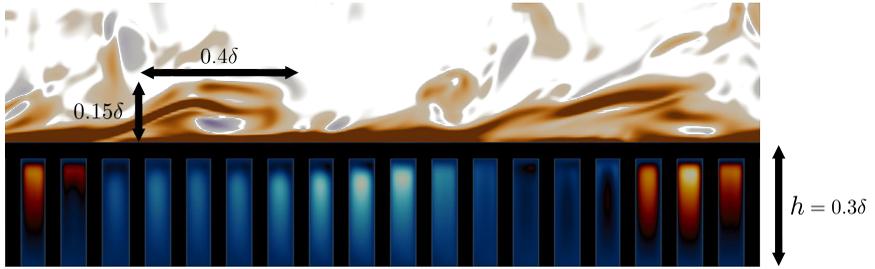

Figure 12: Vortex visualization using the $Q$-criterion for substrate $HP1$ in the $x$-$y$ plane. The vortex core (light color) is a region of $Q > 0$ (high vorticity) and is surrounded by a sheet (dark color) of $Q < 0$ (high shear). The hot and cold regions below the surface represent positive and negative wall-normal velocity fluctuations, respectively. Flow direction is from left to right.

Recalling the drag decomposition in figure 6, it was observed that increased streamwise-preferential anisotropy had a drag reducing effect in the $LP$ cases, albeit a small one. As it is clear now that these cases belong to the smooth-wall-like turbulence regime, this makes the observed drag change in line with the results of Gómez-de Segura & García-Mayoral (2019). In the smooth-wall-like regime, the changes in drag are due to the "virtual-origin" effect (Luchini *et al.* 1991; Jiménez 1994; Luchini 1996; García-Mayoral *et al.* 2019; Ibrahim *et al.* 2021; Habibi Khorasani *et al.* 2022), where the Reynolds stress generating quasi-streamwise vortices become displaced farther away from the surface where a slip velocity is present, resulting in a net drag reduction. This linear mechanism of passive drag reduction is only achievable so long as the flow in the immediate proximity of the surface remains Stokes-like (Luchini 2015), and becomes negated beyond it.

## 4. Surface flow

The focus is now placed on the permeable surface of the substrates at $y^+ = 0$, which is the region directly in contact with the overlying turbulent flow. Figure 13 shows the velocity and pressure fluctuations at $y^+ = 0$; the differences that were observed in the flow-fields of $HP1$ and $LP1$ above the surface (figure 4) are also reflected here. It can be observed from figure 13a that $HP1$ lacks the streaky patterns of $LP1$, shown in figure 13b, but has more intense activity. The imprint of the surface geometry is also more clearly visible in the flow-field of $HP1$. For the wall-normal velocity, spanwise coherent patterns are observable for $HP1$ (figure 13c) whereas such coherency is not discernible for $LP1$ (figure 13d).

The spectral energy densities at the surface level in figure 14 reveal the signature of the K-H-like structures (indicated with the red lines) observed in figure 11 for $HP1$. For $LP1$, no such signature is visible and overall far less flow activity occurs at the permeable surface. Additionally, the spectra show energetic regions at wavelengths equal to the horizontal substrate spacings $s_x^+$ and $s_z^+$ (indicated using the green lines) along with their sub-harmonic wavelengths. These regions represent the pore-coherent flow which is modulated by the ambient turbulence, as similarly occurs over rough surfaces (Abderrahaman-Elena *et al.* 2019). The pore-coherent flow component forming along the spanwise direction repeats periodically along the streamwise direction in intervals of $\lambda_x^+ = s_x^+$. This flow component is modulated by the ambient turbulence and becomes amplified over a broad range of spanwise wavelengths as can be seen in figures 14a and 14b for the *u* and *v* spectra of $HP1$. A similar effect takes place for the pore-coherent flow component forming along the streamwise direction. While the pore-coherent flow does exist for $LP1$, it is significantly weaker compared to $HP1$. Regarding the $HP2'$ and $HP3'$ cases, some delicate differences can be observed compared to $HP2$ and $HP3$, but the assessment of them is not done here and is instead gathered in appendix C for the interested reader.

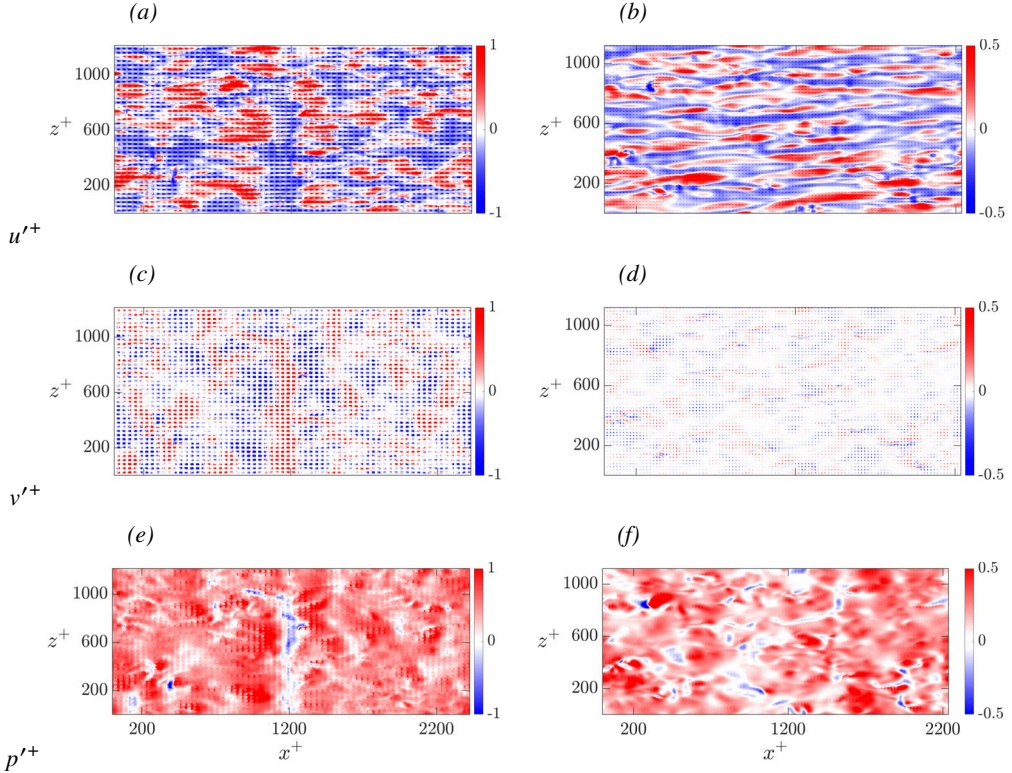

Figure 13: Instantaneous fluctuations of *(a, b)* streamwise velocity, *(c, d)* wall-normal velocity, and *(e, f)* pressure at $y^+ = 0$. First column, $HP1$; second column, $LP1$. Flow direction is from left to right. The white regions are due to the presence of the porous substrate's rods.

Ultimately, the amplified pore-coherent flow (areas enclosed by green lines in the spectra of figure 14) and its sub-harmonics for the $HP$ cases can be attributed to the existence of energetically coherent ambient turbulent scales, particularly those associated with the K-H-type structures. This would explain why the $LP$ cases, despite having streamwise and spanwise pitch-lengths of comparable size to those of the $HP$ cases, do not exhibit similarly strong pore-coherent flows. Broadband excitation of any flow component induced by the geometry of the porous substrate is contingent upon the existence of broadband energetic turbulent structures. The observations made here have important implications for the sub-surface flow since the pore-coherent flow, which undergoes modulation by the ambient turbulence, factors into the scale-selection that takes place at the surface and therefore the scales of motion that occur inside the substrates.

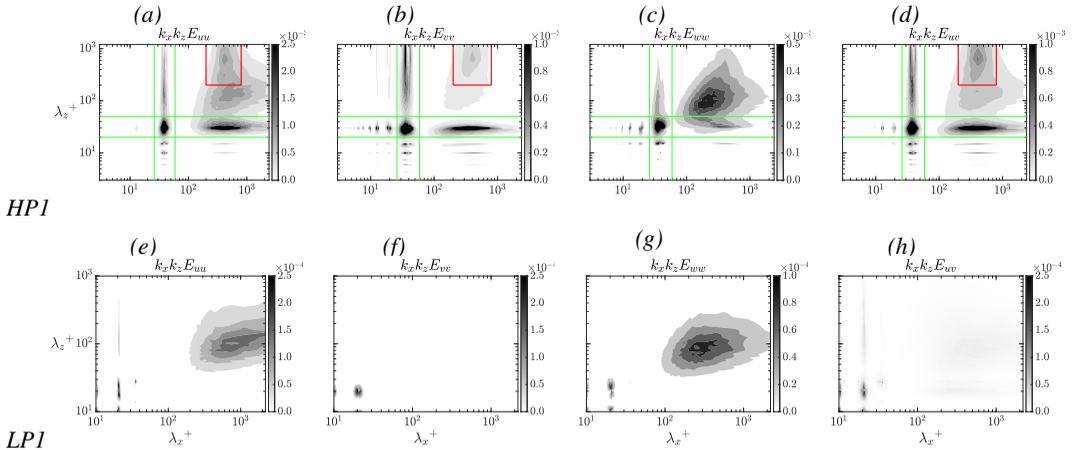

Figure 14: Pre-multiplied two-dimensional spectral densities: (*a, e*) $k_x k_z E_{uu}$, (*b, f*) $k_x k_z E_{vv}$, (*c, g*) $k_x k_z E_{ww}$, and (*g, h*) $k_x k_z E_{uv}$ at $y^+ = 0$. First row, *HP*1; second row, *LP*1. The red lines demarcate $200 \lesssim \lambda_x^+ \lesssim 800$ and $200 \lesssim \lambda_z^+$. The green lines enclose the signature of the pore-coherent flow which are coincident with $\lambda_x^+ = s_x^+$ and $\lambda_z^+ = s_z^+$.

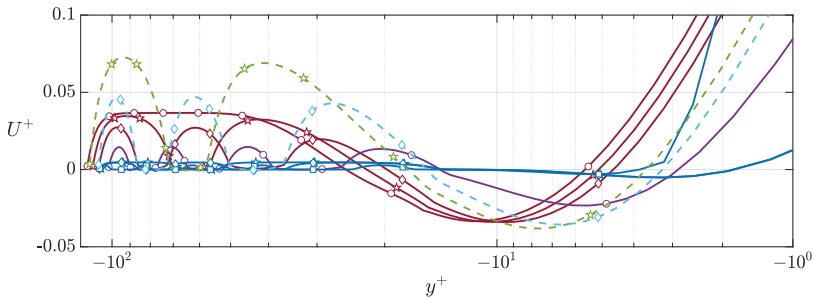

Figure 15: Mean velocity profiles inside the substrates for the cases of table 1.

## 5. Sub-surface flow

Thus far, the effects due to the presence of a porous substrate have been examined for the overlying flow. Attention is now given to the sub-surface flow that develops inside the substrates.

### 5.1. *Mean flow, fluctuating velocities and Reynolds shear stress*

First, the mean velocity along with the fluctuations of the different velocity components are examined. Figure 15 demonstrates that the mean flow develops a shear-layer beneath the surface where the flow decays rapidly over a short distance in an exponential manner. This is mainly notable for the *HP* cases, since a very weak mean flow develops inside the substrates for the *LP* cases. Within the shear-layer, the mean flow exhibits flow reversal. Larger wall-normal pore spacings cause this region to become extended and the position at which the flow undergoes reversal corresponds to the bottom of the first pore layer, i.e. $y^+ \approx -s_y^+$. The exponential decay exhibits similarity across the different $HP$ cases, but seemingly requires both a large enough wall-normal pore spacing and a strong surface flow to develop since the $LP$ cases do not exhibit such a quality.

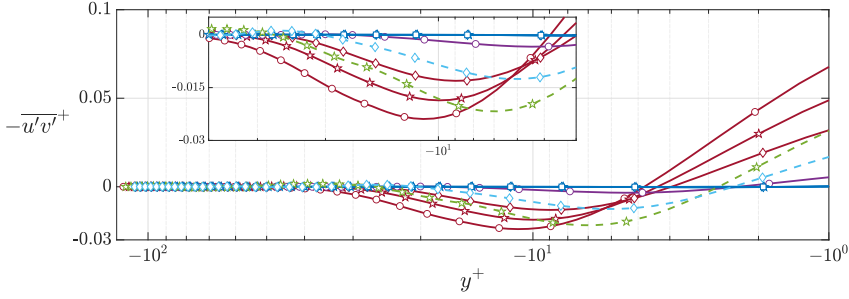

Figure 16: Reynolds shear stress profiles inside the substrates for the cases of table 1.

The Reynolds shear stress undergoes a similar pattern of sign change as the mean flow below the surface (figure 16). The $HP$ cases once again develop a region of rapid change which the $LP$ cases do not demonstrate; the former cases also have notably greater magnitude. As the first pore layer becomes deeper, this reversal region becomes extended. $HP2'$ and $HP3'$ exhibit the same pattern as the other $HP$ cases.

For the velocity fluctuations (figure 17), all of them gradually decay toward the floor of the porous substrates where they become forcibly dampened due to the no-slip condition. However, both $u'$ and $w'$ undergo dampening at the bottom of each pore layer as evident by their oscillatory patterns, whereas $v'$ largely demonstrates a monotonic decay. Considering the downward path from a surface pore-opening; the wall-normal flow entering an opening will not come across any barriers along the way toward the substrate floor since it moves through what is essentially a narrow duct. For wall-parallel flow however, at the bottom of each pore layer the interconnected rods of the substrate's geometry will impede any in-plane motion. The overall magnitude of the spanwise velocity fluctuations is less than those of the streamwise and wall-normal velocity fluctuations, which follows from the spanwise velocity also being the least energetic velocity component at the surface of the substrates. $HP2'$ and $HP3'$ have stronger streamwise fluctuations compared to $HP2$ and $HP3$. This can primarily be attributed to their larger $K_x^+$, resulting in less impedance of streamwise momentum, but may also be attributable to the stronger turbulence at the surface (figure C.2), resulting in a greater modulation of the sub-surface flow, an aspect which will be examined in §6.

Some of the observations made here have been similarly reported for turbulent flows over engineered dense canopies (Sharma & García-Mayoral 2020), such as the gradual decay of the wall-normal fluctuations. Periodic dampening of the fluctuations were not reported for the canopy flows, but this is attributable to the porous substrates having layers of interconnected solid elements whereas the canopy filaments were isolated from one-another and do not place similar restrictions on in-plane fluid motion.

## 5.2. *Flow structure and features*

The surface flow was described in §4 and bearing in mind the observations made there the sub-surface flow is now examined by assessing the instantaneous fluctuations within the first pore layer at $y^+ \approx -15$ in figure 18. The flow fields of both $HP2$ and $HP3$ show spanwise elongated patterns in both their streamwise and wall-normal velocity fluctuations (figures 18a, 18e, 18b, 18f). The pressure fluctuations of both $HP2$ (figure 18i) and $HP3$ (figure 18j) reflect the patterns of their streamwise and wall-normal velocity fluctuations, suggesting that the velocity fluctuations inside the substrates are induced by the pressure fluctuations. Kuwata & Suga (2016) attributed the velocity fluctuations occurring within the porous substrate to the pressure fluctuations caused by the K-H-like instability at the surface. The observations made here seem to agree with this, as the turbulence in the near-surface region of both $HP2$ and $HP3$ falls into the K-H-like regime. The flow field of $HP3'$ (figures 18d, 18h, 18l) is similar to $HP3$ (figures 18b, 18f, 18j) with no discernible differences existing between them. The flow field of $HP2'$ in figures 18c, 18g, 18k shows a stronger spanwise coherency than $HP2$ in figures 18a, 18e, 18i (this greater coherency can also be observed in the flow-field at $y^+ = 0$ in figure C.1, and is attributed to the stronger K-H-like scales visible in the spectra of figure C.2).

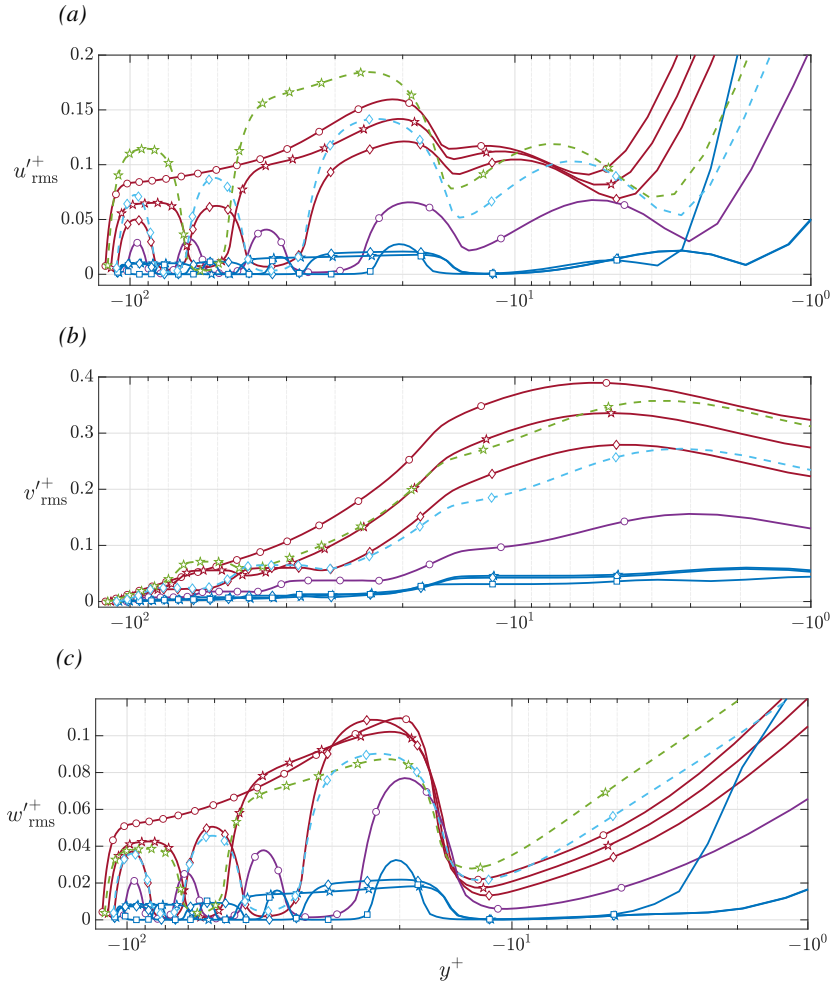

Figure 17: Streamwise (*a*), wall-normal (*b*) and spanwise (*c*) r.m.s. velocity fluctuations within the porous substrates.

More details are revealed by examining the spectra of the fluctuations within the first pore layer at $y^+ = -15$, shown in figure 19. The spectra of $HP2$ and $HP3$ (figures 19a-d and 19e-h) show that almost no ambient turbulence scales penetrate into the substrate. Only the pore-coherent flow remains energetically discernible, as seen by its spectral signature enclosed by the green lines of figure 19. The pore-coherent flow is also diminished compared to the surface region (figure C.2), but not as strongly as the ambient turbulence.

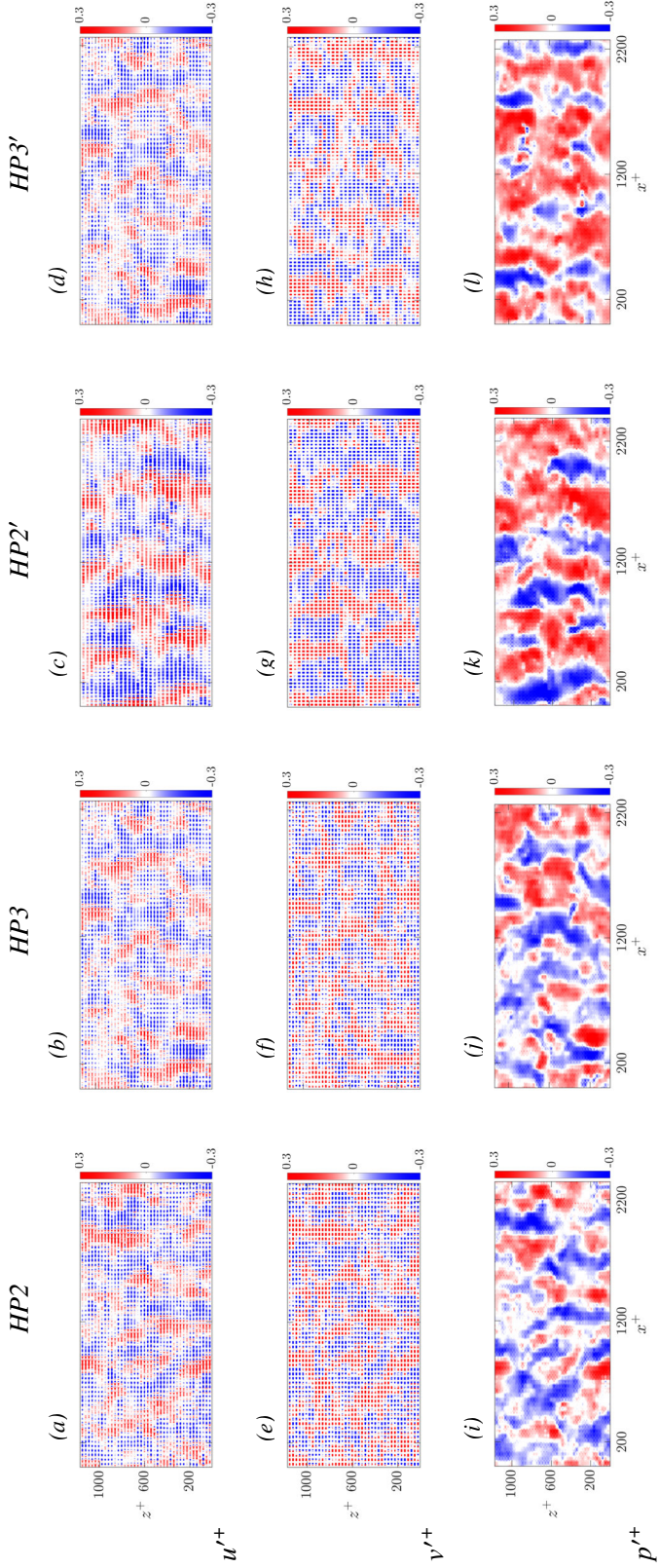

Figure 18: Instantaneous fluctuations of ($a$, $b$, $c$, $d$) streamwise velocity, ($e$, $f$, $g$, $h$) wall-normal velocity and ($i$, $j$, $k$, $l$) pressure at $y^+ \approx -15$. First column, $HP2$; second column, $HP3$; third column, $HP2'$; fourth column, $HP3'$. Flow direction is from left to right. The white regions are due to the presence of the porous substrate's rods.

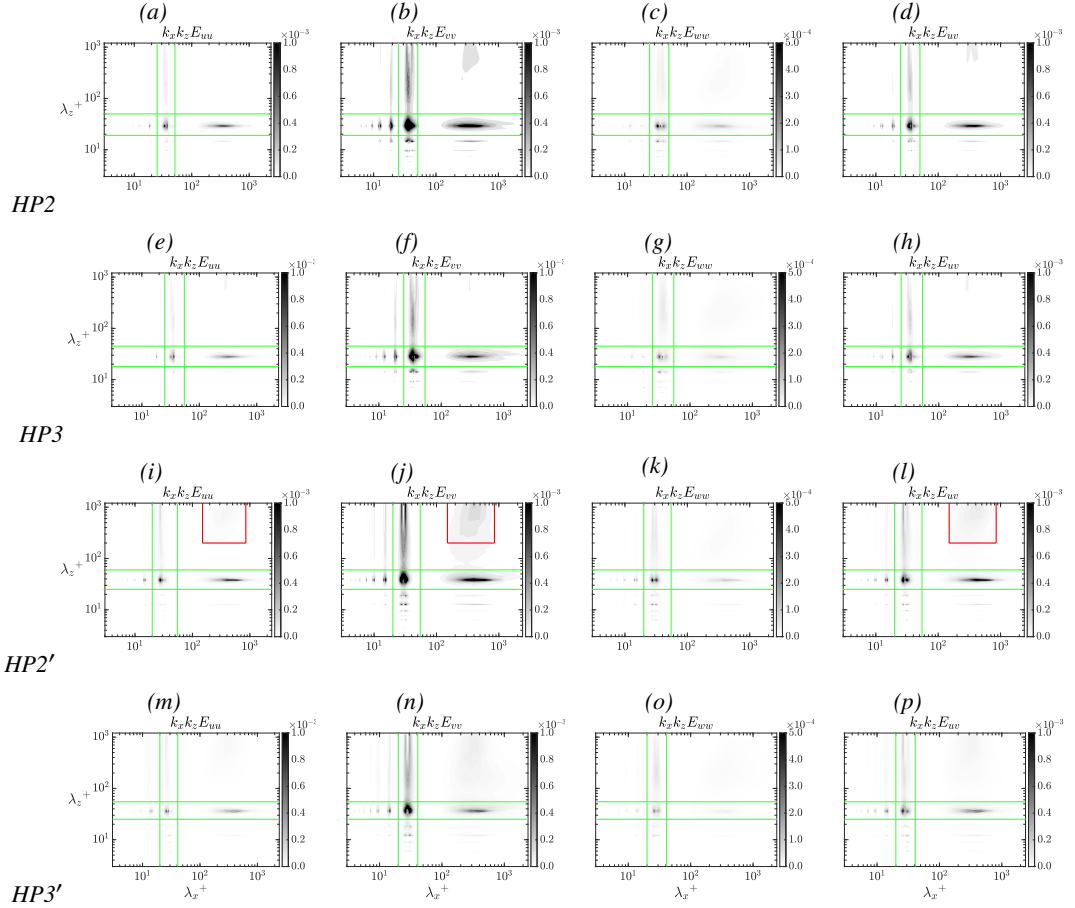

Figure 19: Pre-multiplied spectral densities: (*a, e, i, m*) $k_x k_z E_{uu}$, (*b, f, j, n*) $k_x k_z E_{vv}$, (*c, g, k, o*) $k_x k_z E_{ww}$, and (*d, h, l ,p*) $k_x k_z E_{uv}$ at $y^+ \approx -15$. First row, *HP2*; second row, *HP3*; third row, *HP2′*; fourth row, *HP3′*. Note the differences in the overall magnitude of the contours for the different cases. The green lines enclose the most energetically significant parts of the pore-coherent flow and the red lines those of the K-H-like rollers.

For *HP2′*, its stronger K-H-like scales at the surface level lead to the survival of those scales down to this depth within the substrate (the regions enclosed by the red lines in figures 19i, 19j, 19l), although they are quite weak. Despite having the same wall-normal permeability as *HP2*, the streamwise favorable anisotropy of *HP2′* leads to stronger turbulent scales at the surface which are then able to penetrate deeper into the substrate.

Owing to the fact that turbulence does not survive this deep into the porous substrates for *HP2*, *HP3*, *HP2′* and *HP3′*, the coherent patterns observed in their flow fields (figure 18) must be attributed to the pore-coherent flow which remains detectable at this depth. The broadband spanwise intensification of the pore-coherent flow however is imparted to it from the surface level turbulence which possess spanwise coherent energetic scales. The modulation persists throughout the substrate and hence why the spectra in figure 19 show long patches of spanwise energetic scales, particularly for the wall-normal velocity.

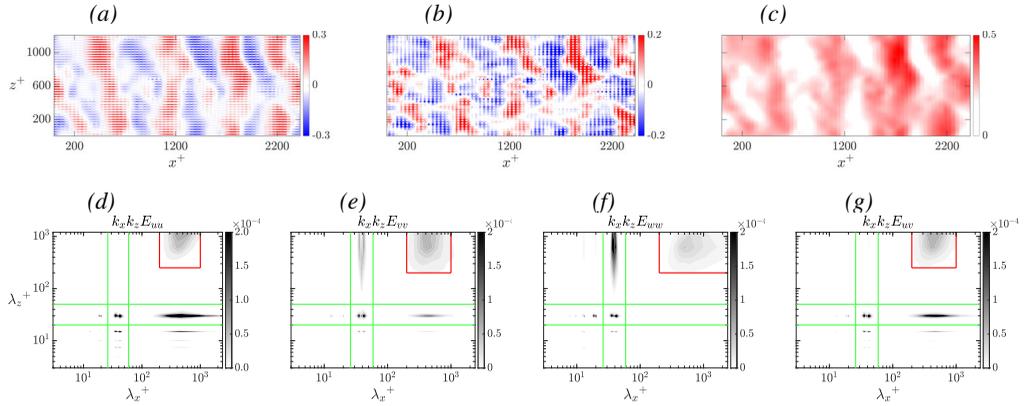

Figure 20: Instantaneous velocity fluctuations and pre-multiplied spectral energy densities of case $HP1$ at $y^+ \approx -55$: (*a*) streamwise fluctuations, (*b*) wall-normal fluctuations, (*c*) pressure fluctuations; (*e*) $k_x k_z E_{uu}$, (*f*) $k_x k_z E_{vv}$, (*g*) $k_x k_z E_{ww}$, and (*h*) $k_x k_z E_{uv}$. The green lines enclose the most energetically significant parts of the pore-coherent flow and the red lines the surviving turbulence belonging to the K-H-like scales.

Ultimately, for the porous substrates under consideration here, there exists a notable pore-coherent flow component below the surface, and in some cases weak scales of ambient turbulence related to the K-H-like instability. Similarly, in flows over canopies the fluctuations below the canopy tip-plane are attributed to the strong overlying cross-flow rollers that develop due to the existence of a perturbed mixing layer (Sharma & García-Mayoral 2020). As mentioned previously in §3.5, for a mixing layer to emerge over porous media very high effective surface permeability (or permeability Reynolds number) is required (Manes *et al.* 2011). Outside of this mixing layer regime, the major flow activity below the surface seems to mainly be due to the pore-coherent flow which undergoes modulation by the turbulence at substrate's surface. Manes *et al.* (2011) examined whether the resulting eddy structures over their porous foams shared the same characteristics as those reported over canopies and which are associated with an inflectional instability of the mean velocity (White & Nepf 2007). They observed this to not be the case for low to intermediate ranges of permeability. This also applies to the porous substrates examined in this paper and the analysis done to quantify this is gathered in appendix B.

Before proceeding further, as a final examination to see whether the flow structure inside the substrates undergoes any notable change deeper inside the substrate, the instantaneous fluctuations as well as the spectra at $y^+ \approx -55$ for case $HP1$ alone are shown in figure 20. One can witness that the patterns are overall similar to those observed at $y^+ \approx -15$ for the rest of the $HP$ cases, with the most notable scales of motion again being those of the pore-coherent flow, while a weak footprint of the K-H-like scales are also present. The existence of the latter at this depth is of course attributable to the stronger overlying K-H-like structures of $HP1$. In addition, $HP1$ also lacks interconnected rod layers inside the substrate which would impede downward directed flow. The explanation for the spanwise coherence of the flow-field is similar to what was previously described for the flow at the shallower depth of $y^+ \approx -15$. The spectral signature of the pore-coherent flow which encompasses a range of streamwise scales is repeated along the spanwise direction at intervals equal to the spanwise spacing ($\lambda_z^+ = s_z^+$), i.e. between two consecutive pores along this direction. In essence, the patches are collections of narrow fingers of streamwise velocity which are confined to the pores due to a micro-channelization effect. These flow elements appear as a spanwise coherent region macroscopically due to being modulated in amplitude, which is what will be examined next.

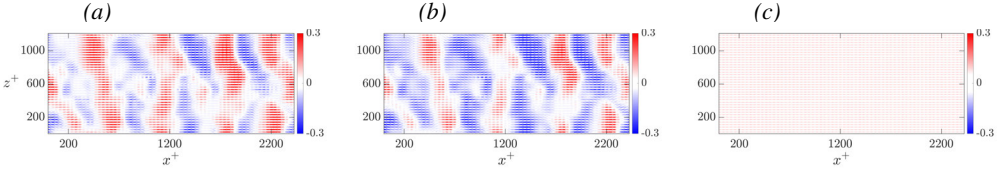

Figure 21: Decomposition of fluctuating streamwise velocity of case $HP1$ at $y^+ \approx -55$ according to (6.2): $(a)$ $\boldsymbol{u'}$, $(b)$ $\boldsymbol{u''}$, $(c)$ $\boldsymbol{\tilde{u}}$.

## 6. Surface-flow induced amplitude modulation of sub-surface flow

The pore-coherent flow is subject to amplitude modulation (AM) by the scales of the overlying ambient turbulence, evidence of which was provided in the experimental investigation of Kim, Blois, Best & Christensen (2020). This phenomena will now be examined for the substrates considered in this study.

The presence of any solid structure introduces spatial inhomogeneities within the flow field. A conventional method for isolating this effect is the triple decomposition of Reynolds & Hussain (1972)

$$\boldsymbol{u} = \boldsymbol{U} + \boldsymbol{u'}, \tag{6.1}$$

$$\boldsymbol{u'} = \boldsymbol{\tilde{u}} + \boldsymbol{u''}, \tag{6.2}$$

where $\boldsymbol{u}$ is the total velocity, $\boldsymbol{U}$ the time- and space-averaged mean velocity and $\boldsymbol{u'}$ the fluctuating velocity. The fluctuating component itself then consists of a spatially inhomogenous time-averaged component, $\boldsymbol{\tilde{u}}$, called the dispersive velocity, and a turbulent component, $\boldsymbol{u''}$. Amplitude modulation is a dynamic effect that is not reflected in the time-averaged dispersive velocity field. The fluctuating velocity along with its different components from (6.2) are shown in figure 21 for case $HP1$. The time-averaged dispersive velocity field is weak and does not have irregularities, but as is evident in the spectra of figure 20d, the pore-coherent flow which resides in $\boldsymbol{u''}$ has a different spatial pattern.

Abderrahaman-Elena *et al.* (2019) proposed a modified triple decomposition and used it to quantify the AM effect for rough-wall turbulence. However, for the porous substrates examined here the wavelengths of the pore-coherent flow and those of the ambient turbulence do not significantly overlap and regular Fourier filtering can be used to isolate this effect. This is demonstrated in figure 22, where the high-frequency (low-wavelength) amplitude-modulated signal of the pore-coherent flow has been removed from the streamwise velocity signal at the surface ($y^+ = 0$) using low-pass filtering. This recovers the low-frequency (long-wavelength) signal of the ambient turbulence. There does not seem to be a discernible AM effect above the surface at $y^+ \approx 5$, as the pore-flow component (the undulations of the black line) do not undergo notable changes in amplitude. The AM effect is similarly demonstrated for the wall-normal velocity in figure 23. Note that this AM phenomena is different from AM observed in canonical turbulent flows between the inner and outer flow regions (Mathis, Hutchins & Marusic 2009). That effect is due to the existence of large-scale structures within the log-layer which will emerge when the Reynolds number is sufficiently large ($Re_\tau > 1700$).

The approach undertaken here to quantify AM follows that of Mathis, Hutchins & Marusic (2009), where the correlation between the low-pass filtered (large-scale) streamwise velocity fluctuations, $u'_L$, and the long-wavelength envelope of the high-pass filtered (small-scale) velocity fluctuations, $E_L(\boldsymbol{u'}_s)$, taken at two different fixed $y$ positions ($y_1$ for the $u'_L$ and $y_2$ for $u'_s$) quantifies the degree of AM (note that $\boldsymbol{u'}_s$ can be any of the velocity components)

$$R_{\boldsymbol{u}}(y_1, y_2) = \frac{\overline{u'_L \, E_L(\boldsymbol{u'}_s)}}{\sqrt{\overline{u'_L}^2}\sqrt{\overline{E_L(\boldsymbol{u'}_s)}^2}}. \tag{6.3}$$

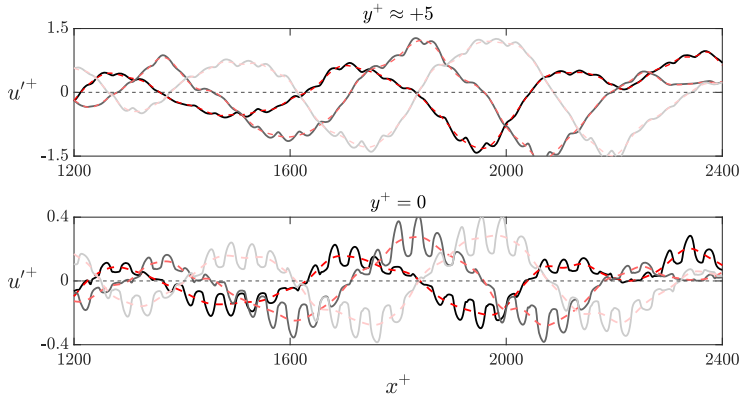

Figure 22: Raw (——) and low-pass filtered (— —) spanwise-averaged streamwise velocity fluctuations of case $HP1$ at $y^+ \approx 5$ and $y^+ = 0$, respectively. The solid to faded lines represent three successive time signals.

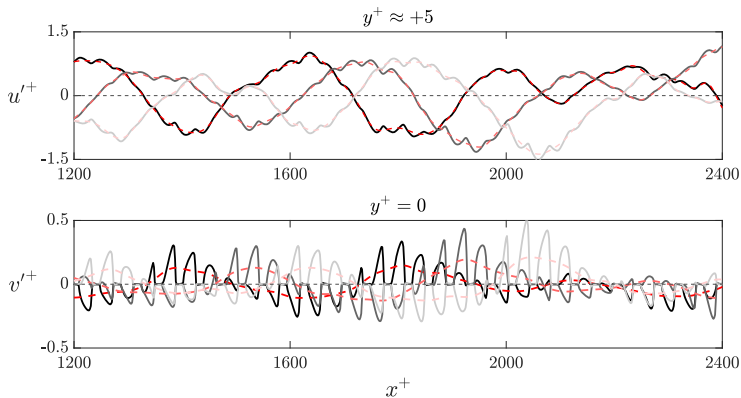

Figure 23: Raw (——) and low-pass filtered (— —) spanwise-averaged streamwise velocity fluctuations at $y^+ \approx 5$, and the spanwise-averaged wall-normal velocity fluctuations at $y^+ = 0$ for $HP1$. The solid to faded lines represent three successive time signals.

In (6.3), $E$ denotes the envelope of a signal and is acquired using the Hilbert transform. The Hilbert transform of a real-valued function, $f(t)$, produces another real-valued function, $\tilde{f}(t)$. Together, $f(t)$ and $\tilde{f}(t)$ form a harmonic conjugate pair and define the complex analytic signal of $f(t)$,

$$F(t) = f(t) + i\tilde{f}(t) = E(t)e^{i\phi(t)}. \tag{6.4}$$

This provides the instantaneous envelope, $E(t)$, and phase, $\phi(t)$, allowing for the demodulation of the original modulated signal, $f(t)$. More details regarding the Hilbert transform may be found in Mathis *et al.* (2009) and the references contained therein. Other approaches can also be used to the assess modulation effects, such as wavelet analysis used by Baars *et al.* (2015) to quantify AM and FM effects. The Hilbert-based approach however remains robust. The long-wavelength envelope of small-scale velocity, $E_L(\boldsymbol{u'_s})$, obtained after taking the Hilbert transform of the velocity time-signal, is then high-pass filtered to keep only the modulated small-scale velocity signal. Unlike in experimental measurements, the velocity signals here are not single-point measurements. Instead, spanwise-averaged one-dimensional velocity signals at different $y^+$ planes are used to first obtain instantaneous correlations and then followed by ensemble-averaging over all temporal samples to obtain a single correlation coefficient.

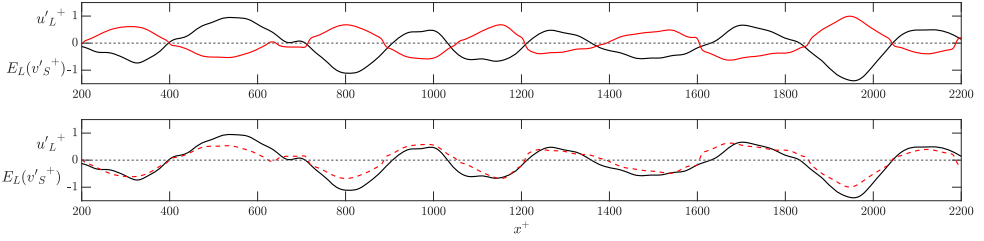

Figure 24: Amplitude modulation in action. The top plot shows large-scale streamwise velocity fluctuations (—) at $y^+ \approx +5$ and the long-wavelength envelope of small-scale wall-normal velocity fluctuations (—) at $y^+ = 0$ of case $HP2$; the bottom figure shows the same, only with the envelope (– –) now phase-shifted by $\pi$.

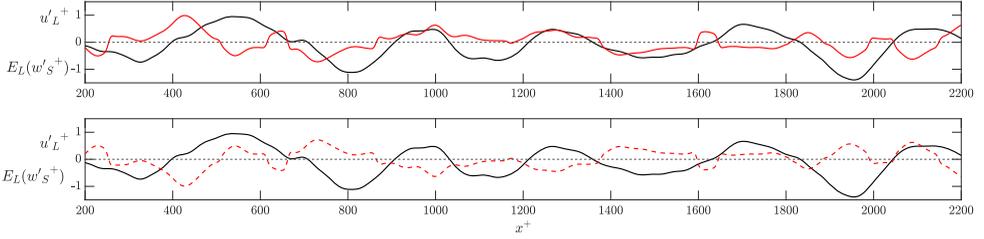

Figure 25: Lack of amplitude modulation. The plots are similar to those in figure 24, but with the envelope of the spanwise velocity instead of the wall-normal velocity.

The plot at the top of figure 24 shows how the small-scale wall-normal velocity fluctuations at the permeable surface are modulated in amplitude by the large-scale streamwise velocity fluctuations of the ambient turbulence above it. The envelope of $v'_s$ rises and falls along with the variations in the amplitude of $u'_L$. The high degree of correlation becomes clearer when considering the bottom plot, where the envelope of $v'_s$ is phase-shifted by $\pi$, making it overlap to a significant extent with the signal of $u'_L$. This also demonstrates that events of $u$ and $v$ are almost always in anti-phase with respect to one another close to the surface. Such a modulation effect is not observed between $u'_L$ and $w'_s$ in figure 25. The lack of AM from the outer to inner region is also demonstrated using this approach in appendix D.

The instantaneous frequencies of the velocities can be calculated from their analytic signals obtained using the Hilbert transform. These can then be used to calculate the instantaneous phase-difference between the streamwise and wall-normal velocities. A probability density histogram of instantaneous phase differences is shown in figure 26b for case $HP1$, demonstrating that $u'$ and $v'$ are predominately in anti-phase. The probability density histogram for the AM correlation coefficient (6.3) is shown in figure 26a and demonstrates the persistent presence of AM between the $u'$ and $v'$ as they evolve in time. Similar effects are not observed between $u'$ and $w'$ in figures 26d and 26c.

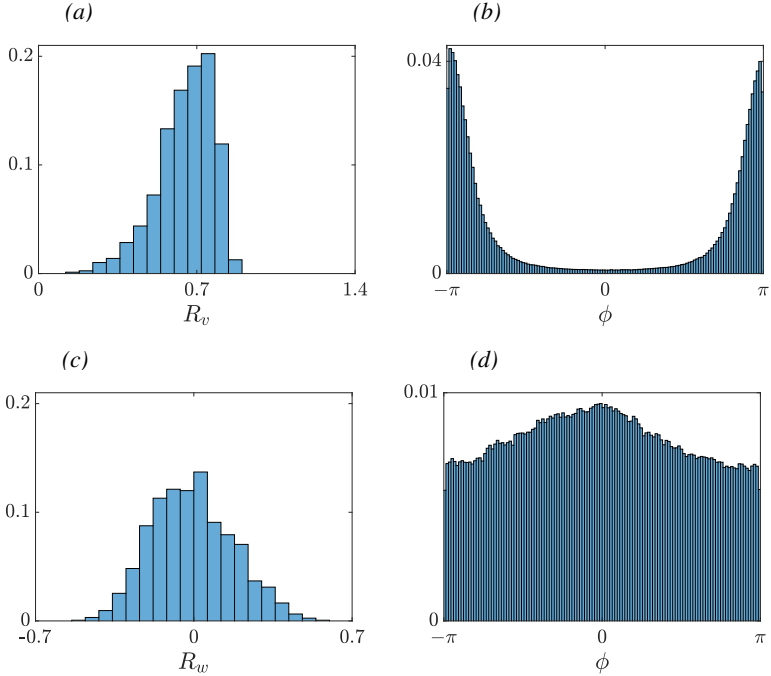

Figure 26: Probability density histograms of AM correlation between (*a*) $u'_L{}^+$ and $E_L(v'_s{}^+)$, and (*c*) $u'_L{}^+$ and $E_L(w'_s{}^+)$. Instantaneous phase differences between (*b*) $u'_L{}^+$ and $v'_L{}^+$, (*d*) $u'_L{}^+$ and $w'_L{}^+$ for case *HP*2. The $u'$ signal was taken at $y^+ \approx 5$ while the $v'$ and $w'$ were taken at $y^+ = 0$.

Examination of the flow inside the substrates (figures 20 and 21) demonstrated that AM is present within them. As such, it is of interest to see how deep the effect persists and also how its strength differs between the various substrates. Figure 27 displays the AM effect on the wall-normal velocity at different depths for the substrates of table 1. AM remains quite strong half-way down into the substrates for the *HP* cases. It is much weaker overall for the *LP* cases, highlighting how the energetic surface level dynamics of the *HP* cases which fall into the K-H-like regime enhances this effect.

## 7. Summary and discussion

Direct numerical simulations of turbulent flows in an open channel geometry where the wall-side of the channel is covered by a porous substrate have been carried out in this work. Anisotropic porous substrates with permeability components of different values were first assessed in how they cause changes in the overlying turbulent flow. When the wall-impedance condition becomes weakened, near-wall turbulence undergoes a transition away from its canonical structure –characterized by the presence of streaks and quasi-streamwise vortices– to one where spanwise coherent structures reminiscent of the K-H instability emerge.

The primary permeability component of significance in determining wall-impedance is $K_y$. An analysis using the $K_{Br}$ permeability criteria of Sharma *et al.* (2017) and Gómez-de Segura *et al.* (2018) for predicting when turbulence transitions to a K-H-like regime remains robust for the DNS data in this study. The $K_{Br}$ condition was obtained using linear stability analysis of permeable wall boundary conditions derived using the Darcy-Brinkman equation but remain applicable to the pore-scale resolved DNS data in this study, indicating that the microstructure details of the porous substrates do not have a leading-order impact on the instability. This agrees with the argument made by White & Nepf (2007), who assessed that only the overall resistance of the porous layer is important and not the details of its geometry in bringing about and sustaining the instability.

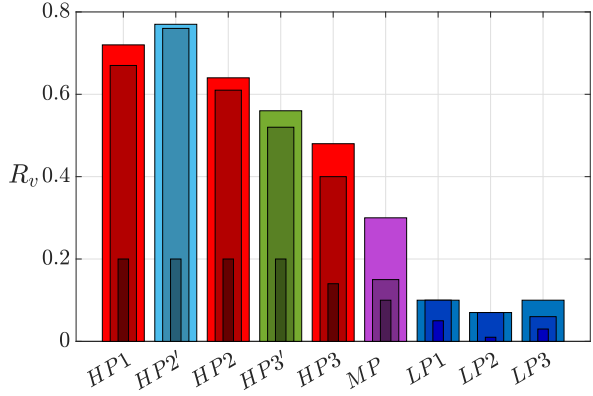

Figure 27: Degree of amplitude modulation for the different substrates from the surface adjacent region down to the impenetrable floor of the substrates. *(a)* Modulation for wall-normal velocity, and *(b)* modulation for spanwise velocity. Light to darker shades correspond to $y^+ \approx -5$, $y^+ \approx -50$ and $y^+ \approx -110$, respectively.

Past results in the literature using continuum-based approaches of representing a porous region or using permeable wall boundary conditions have suggested that a reduction in drag is perhaps attainable for certain combinations of permeability, particularly for streamwise preferential anisotropy (Gómez-de Segura & García-Mayoral 2019). However, none of the porous substrates examined here resulted in drag reduction (figure 5a). Drag reduction in a passive manner can be obtained if a surface can simultaneously weaken viscous dissipation while impeding turbulent mixing from taking place close to its vicinity, an effect which is quantified in the "virtual-origin" framework (Luchini 1996; Ibrahim *et al.* 2021). The weakening of viscous dissipation is typically quantified in terms of a slip velocity, which is negligible for the porous substrates tested in this study. Turbulent activity however increases in the vicinity of the surface such that the net effect becomes one of drag increase (figure 6). Streamwise-preferential anisotropy does not lead to a drag reducing effect, which can only occur in the canonical smooth-wall-like regime of near-wall turbulence, and which only the *LP* cases belong to. They do however exhibit a trend of drag decrease with increased streamwise-preferential anisotropy. Ultimately, drag reduction using porous media must be assessed in terms of the slip-velocity (or slip-length) they can cause at the surface.

At the surface of the porous media, spectral analysis reveals the existence of flow signatures conforming to the geometry of the surface and with amplified levels of energy (figure 14). Inside the porous substrates, the surviving turbulence scales become rapidly dampened and the flow component with significant energy-wise is the pore-coherent flow figure 19. The structure of the pore-coherent flow is geometry dependent, which makes the microstructure of the porous medium, particularly at the surface, an important aspect of its design.

The aforementioned pore-coherent flow undergoes significant amplitude modulation by the ambient turbulent motion present near the surface of the porous media. This AM effect extends deep into the porous media, perturbing its flow and becoming a principal means of inducing flow activity inside it. Stronger ambient turbulence at the surface strengthens this effect, such that it becomes more pronounced for the cases which fall into the K-H-like regime figure 26. This is because the K-H-like structures lead to greater momentum exchange between the surface and sub-surface flow. These flow features are conceptually illustrated in figure 28, where going from left to right indicates an increase in $K_y^+$ and ultimately $K_{Br}^+$.

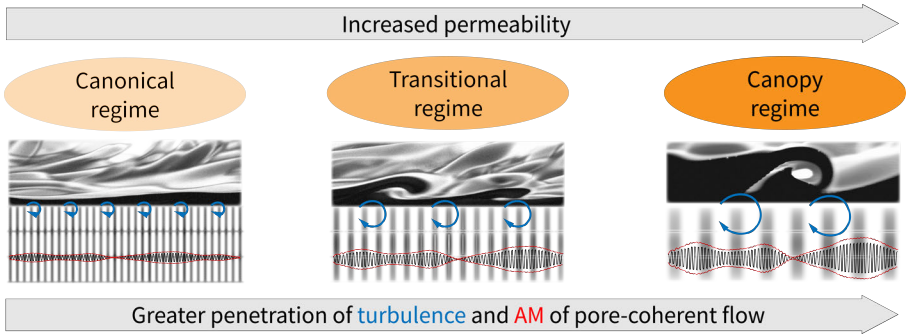

Figure 28: Conceptual schematic showing the evolution of turbulence over porous substrates and the resulting flow phenomena.

Knowledge of the regimes illustrated in figure 28 and the scale interaction which occurs between the porous media and turbulent flow can be leveraged in applications involving heat and mass transfer. Unlike flow momentum, heat transfer stands to benefit from more intense turbulent activity in the vicinity of the porous medium, as this will lead to greater thermal convection. To what degree this can be exploited, is one example of an interesting line of inquiry that can be pursued in relation to turbulence and porous media.

## Acknowledgments

This work was supported by grant SSF-FFL15-0001 from the Swedish Foundation for Strategic Research (SSF) and by the Air Force Office of Scientific Research (AFOSR) grant A9550-19-1-7027 (Program Managers Dr. Gregg Abate and Dr. Douglas Smith). Access to the computational resources at the PDC Center for High Performance Computing, National Supercomputer Centre (NSC) and High Performance Computing Center North (HPC2N) computing centers used for this work were provided by the Swedish National Infrastructure for Computing (SNIC).

## Declaration of Interests

The authors report no conflicts of interest.

# Appendix A. Grid resolution assessment

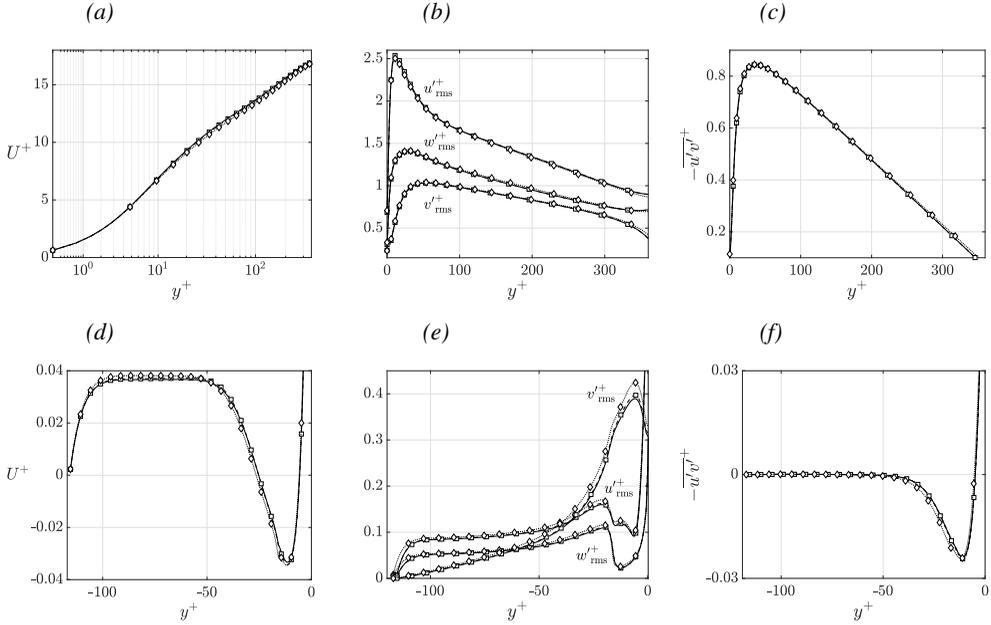

Figure A.1: $(a,d)$ Mean velocity, $(b,d)$ root mean square velocity fluctuations, and $(c,f)$ Reynolds shear stress of the $(a,b,c)$ bulk flow and $(c,d,e)$ substrate regions for $HP1$. The solid lines represent the results from using $d/\Delta_{x,z} = 20$, the symbols –□–, $d/\Delta_{x,z} = 10$; and $\cdots\diamond\cdots$, $d/\Delta_{x,z} = 5$.

| Case | $\sqrt{K_{x_o}}^+$ | $\sqrt{K_{y_o}}^+$ | $\sqrt{K_{z_o}}^+$ | $\sqrt{K_{x_o}}^+_{HR}$ | $\sqrt{K_{y_o}}^+_{HR}$ | $\sqrt{K_{z_o}}^+_{HR}$ | $\Delta_x\%$ | $\Delta_y\%$ | $\Delta_z\%$ |
|---|---|---|---|---|---|---|---|---|---|
| $HP1$ | 3.207 | 3.437 | 4.721 | 3.189 | 3.417 | 4.694 | $-0.56$ | $-0.59$ | $-0.58$ |
| $HP2$ | 2.738 | 2.623 | 3.844 | 2.723 | 2.608 | 3.823 | $-0.55$ | $-0.58$ | $-0.55$ |
| $HP3$ | 2.191 | 2.194 | 2.805 | 2.180 | 2.182 | 2.791 | $-0.51$ | $-0.55$ | $-0.50$ |
| $MP$ | 1.495 | 1.534 | 1.495 | 1.487 | 1.525 | 1.487 | $-0.54$ | $-0.59$ | $-0.54$ |
| $LP1$ | 1.038 | 0.734 | 1.038 | 1.032 | 0.729 | 1.032 | $-0.58$ | $-0.69$ | $-0.58$ |
| $LP2$ | 0.912 | 0.623 | 0.912 | 0.907 | 0.619 | 0.907 | $-0.55$ | $-0.65$ | $-0.55$ |
| $LP3$ | 0.503 | 0.503 | 0.503 | 0.499 | 0.499 | 0.499 | $-0.80$ | $-0.80$ | $-0.80$ |

Table A.1: Darcy permeability estimates for the substrates of table 1 using Stokes flow simulations of REVs (such as those shown in figure 3). The permeabilities with the subscript $HR$ have a grid resolution of $d/\Delta_{x,z} = 20$ and those without $d/\Delta_{x,z} = 10$. The last three columns list the differences in the permeabilities obtained using the two resolutions.

While the resolution requirements for regular channel flow simulations are well-established throughout the literature (Kim & Moin 1985; Lee & Moser 2015), problems involving fluid-solid interactions need to be assessed on a case-by-case basis. The baseline grid has a resolution of $(N_x, N_y, N_z) = (1620, 324, 810)$. This grid more than suffices for resolving the bulk flow region, but it must be determined whether or not the wall-parallel resolution is sufficient for resolving the solid phase of the porous substrates. With this baseline configuration, the number of wall-parallel grid-points per substrate rod thickness becomes $d/\Delta_{x,z} = 10$. This was chosen based on the grid study results of Sharma & García-Mayoral (2020) which was conducted for turbulent flows over canopies. They showed that such a concentration of points per canopy element was enough to resolve them and their induced flow. Nevertheless, to ensure that this resolution is sufficient, a grid study for case $HP1$ was carried out using both a coarser and refined grid. The refined grid had a resolution of $(N_x, N_y, N_z) = (3240, 324, 1620)$ giving $d/\Delta_{x,z} = 20$ while the coarser grid of $(N_x, N_y, N_z) = (810, 324, 405)$ gave $d/\Delta_{x,z} = 5$. The wall-normal grid is kept the same, which is stretched in the bulk flow region and achieves a constant spacing at $y \approx 0.0042$ ($y^+ \approx 15$) resulting in $d/\Delta_y = 25$. Case $HP1$ is chosen since it is has some the strongest flow activity in the substrate region and serves as an appropriate candidate for grid resolution assessment of the $HP$ cases. The $LP$ cases are close to the smooth-wall limit and have very little flow penetrating into the substrate, hence a separate grid resolution assessment for them would be redundant.

From figure A.1, it is clear that the baseline grid with $d/\Delta_{x,z} = 10$ resolves the flow in both the bulk and substrate regions well with the results being grid independent at this resolution. For the coarse grid with $d/\Delta_{x,z} = 5$, the discrepancy is highest immediately below the surface for the r.m.s. wall-normal velocity fluctuations where the error is $\approx 7\%$. The sufficiency of the baseline grid resolution is further reinforced by the results in table A.1, where the calculated Darcy permeabilities for the porous substrates using the baseline and refined grid resolutions is reported. No appreciable improvement in the permeability estimates was achieved by increasing the number of grid points per rod from $d/\Delta_{x,z} = 10$ to $d/\Delta_{x,z} = 20$.

## Appendix B. Assessment of whether or not the K-H-like structures are similar to those of mixing-layer type flows

In the experiments of turbulent flows over porous metal foams conducted by Manes *et al.* (2011), they examined whether the resulting turbulent flow at the permeable surface of the media were similar to those reported by White & Nepf (2007) for sparse porous arrays of cylinders serving as models for vegetation canopies. White & Nepf (2007) demonstrated that the coherent motion at the top of the arrays exhibited a single dominant frequency which was the same as that in free shear layers subject to Kelvin-Helmholtz instability. They therefore characterized the flows as being similar to mixing layers where the instability originates at the inflection point of the mean velocity profile inside the porous region. Manes *et al.* (2011) performed a similar frequency analysis, but observed that only for their highest permeability case did the frequency approach the value associated with mixing-layer type flows. They concluded that for low to moderate permeabilities, the coherent turbulent motion in the vicinity of the permeable surface is not due to the inflectional instability of the mean velocity profile. In a recent DNS study by Wang, Lozano-Durán, Helmig & Chu (2022) where transfer entropy was used to measure causal interactions between porous media and turbulent flows, the frequencies at which these interaction took place over the various porous media were in agreement with the results of Manes *et al.* (2011).

The dominant frequency reported by White & Nepf (2007) was $f = 0.032 \overline{U}/\theta$, where $\theta = \int_{-\infty}^{\infty} (1/4 + (U - \overline{U}/\Delta U)^2) dy$; $\Delta U = U_{y=\delta} - U_P$; $\overline{U} = (U_{y=\delta} + U_P)/2$ and $U_P$ is the velocity deep inside the porous medium. The frequencies reported by Manes *et al.* (2011) for their porous media ($1.9 \leqslant \sqrt{K^+} \leqslant 8.4$) were much higher than 0.032 and only the case of $\sqrt{K^+} = 17.2$ had a frequency close to $f = 0.032 \overline{U}/\theta$.

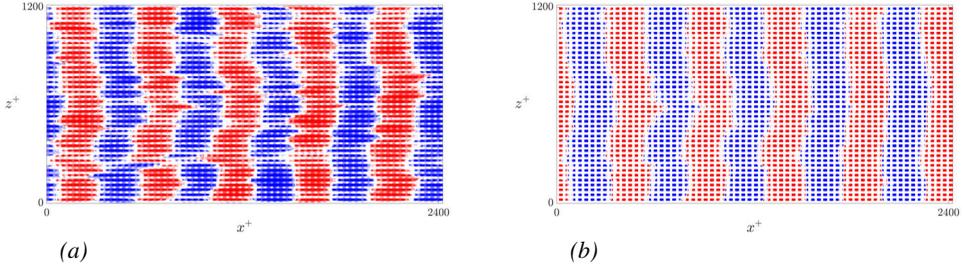

Figure B.1: First SPOD mode at $f = 0.22\overline{U}/\theta$ of the *(a)* streamwise and *(b)* wall-normal velocity fluctuations at the surface ($y^+ = 0$) of $HP1$.

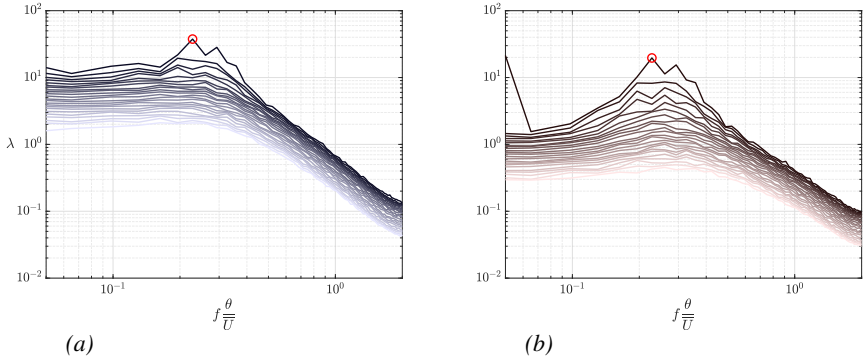

Figure B.2: SPOD eigenvalues of the *(a)* streamwise and *(b)* wall-normal velocity fluctuations at the surface ($y^+ = 0$) of $HP1$. The peak in the leading SPOD mode at $f = 0.22\overline{U}/\theta$ is circled in red. The shading of the curves varies from dark to bright as the mode number increases.

A similar frequency characterization was performed for $HP1$ which has the highest wall-normal permeability $\sqrt{K_y^+} = 3.4$ out of all the cases in table 1. First, spectral proper orthogonal decomposition (SPOD) (Towne *et al.* 2018) is applied to the surface flow of $HP1$. Examination of the SPOD eigenvalues in figure B.2 reveals a peak in the leading SPOD mode at $f = 0.22\overline{U}/\theta$ for both the streamwise and wall-normal velocity components. This is close to $f \approx 0.22\overline{U}/\theta$ reported by Manes *et al.* (2011) for their metal foam which had a permeability of $\sqrt{K^+} = 3.2$. Observing the first SPOD mode for both the streamwise and wall-normal velocity in figure B.1 reveals recurrent spanwise-elongated patterns. Such patterns are not recovered in the SPOD modes for any of the $LP$ cases (not shown). This further demonstrates the regime distinction that was described in §3.4 and §3.5, and it can be concluded that for the porous substrates of table 1, the mixing-layer analogy does not hold, in agreement with the conclusion made by Manes *et al.* (2011).

## Appendix C. Comparison of surface flow features between $HP2$, $HP3$ and $HP2'$, $HP3'$

Thus far, it has been shown that the flow structure close to the surface of the substrates are quite different between the $LP$ and $HP$ cases. The $LP$ cases are essentially smooth-wall like and permit very little turbulent activity from taking place in the proximity of the surface and hence do not motivate further examination. The question now posed is how significant is the pore-coherent flow –determined by wavelengths of the porous medium– relative to the existence of the K-H-like structures. For this purpose, $HP2'$ and $HP3'$ are examined. They are similar to $HP2$ and $HP3$ but with $s_x$ and $s_z$ exchanged (table 1). The wall-normal permeability in this way remains the same but the pore-coherent flow will change and its effect on the flow can be scrutinized. Additionally, the $HP2'$ and $HP3'$ have streamwise preferential anisotropy which will be factored into the analysis that follows.

When considering the flow-fields of $HP2$ and $HP2'$, one can observe a greater degree of spanwise coherence for $HP2'$ in both its streamwise velocity field (figure C.1c) and its wall-normal velocity field (figure C.1g) compared to those of $HP2$ (figures C.1a and C.1g). The spatial patterns of the velocity and pressure fields for $HP2'$ (figures C.1c, C.1g, C.1k) also visibly mirror one another, whereas this is less evident for $HP2$ (figures C.1a, C.1e, C.1i).

Examining the spectra in figure C.2, it can be observed that the ambient turbulence has broadband spanwise coherent scales, in the range of $200 \lesssim \lambda_x{}^+ \lesssim 800$ and $200 \lesssim \lambda_z{}^+$, in both the streamwise velocity (figures C.2i and C.2m), wall-normal velocity (figures C.2j and C.2n) and Reynolds shear stress (figures C.2l and C.2p) spectra of $HP2'$ and $HP3'$. The structure of the ambient turbulence in $HP2$ and $HP3$ is similar. The spectra of $HP2'$ do however demonstrate a patch of more energetic broadband spanwise scales compared to $HP2$ (this could be the reason why the flow-field of $HP2'$ in figure C.1 manifests a greater degree of spanwise coherency compared to $HP2$). When examining the pore-coherent flow, which are the areas enclosed by green lines in the spectra of figure C.2, the broadband streamwise component of them overlap with the ambient turbulence scales to a certain extent such that a degree of scale interaction is taking place between them, but it is not over a wide enough range of scales to alter the structure of the ambient turbulence. For $HP2'$ and $HP3'$, the spanwise pitch length, $s_z{}^+$, is larger compared to $HP2$ and $HP3$, and thus the streamwise pore-coherent flow component overlaps with the ambient turbulence to a greater extent, but still not great enough to be disruptive to the overall dynamics of the turbulence in the near-surface region.

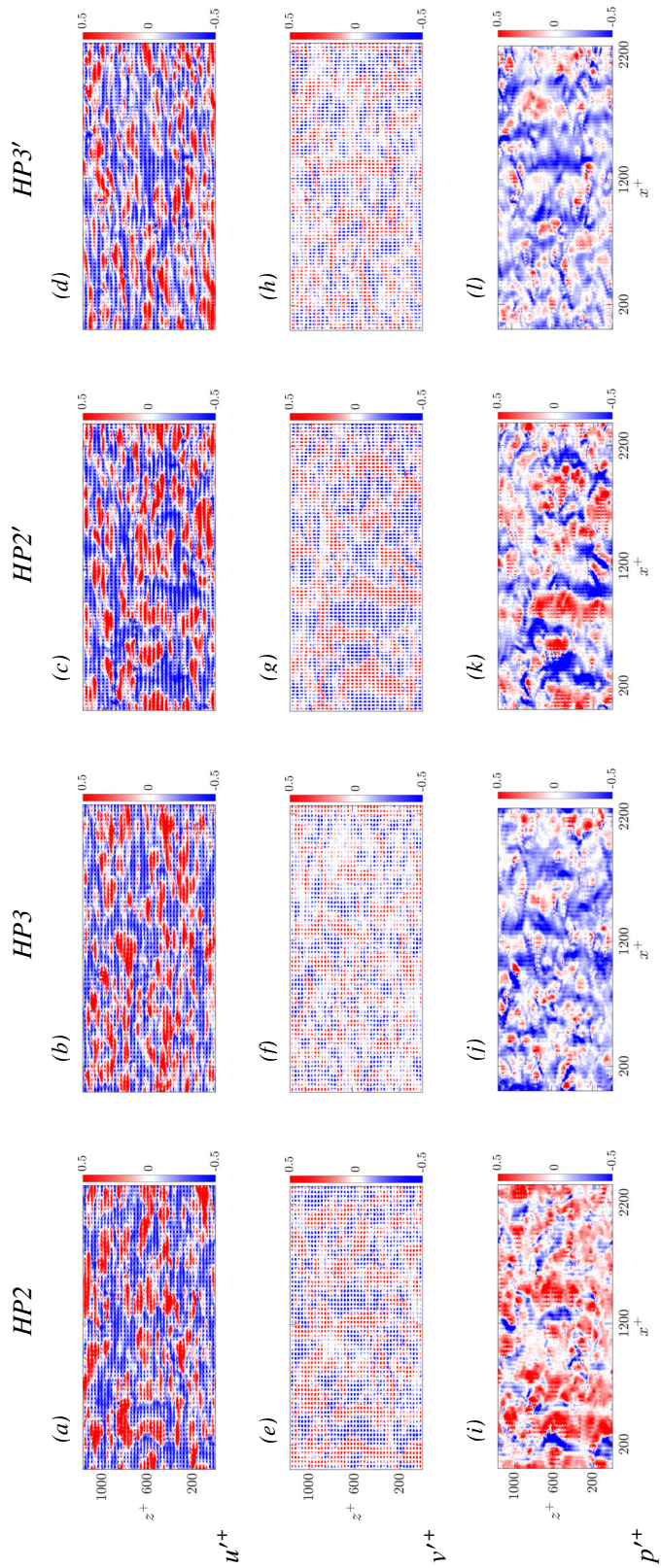

Figure C.1: Instantaneous fluctuations of the ($a$, $b$, $c$, $d$) streamwise velocity, ($e$, $f$, $g$, $h$) wall-normal velocity and ($i$, $j$, $k$, $l$) pressure at $y^+ = 0$. First column, $HP2$; second column, $HP3$; third column, $HP2'$; fourth column, $HP3'$. Flow direction is from left to right.

Figure C.2: Pre-multiplied two-dimensional spectral densities: (*a, e, i, m*) $k_x k_z E_{uu}$, (*b, f, j, n*) $k_x k_z E_{vv}$, (*c, g, k, o*) $k_x k_z E_{ww}$, and (*d, h, l, p*) $k_x k_z E_{uv}$ at $y^+ = 0$. First row, case *HP2*; second row, case *HP3*; third row, case *HP2'*; fourth row, case *HP3'*. The green lines enclose the most energetically significant parts of the pore-coherent flow and the red lines those of the K-H-like rollers.

## Appendix D. AM analysis demonstrating the lack of such an effect imparted from the outer to the inner flow region

Figure D.1: The top plot shows large-scale streamwise velocity fluctuations (——) at $y^+ \approx +100$ and the long-wavelength envelope of small-scale streamwise velocity fluctuations (——) at $y^+ = 0$ of case *HP*1; the bottom figure shows the same, only with the envelope (— —) now phase-shifted by $\pi$.

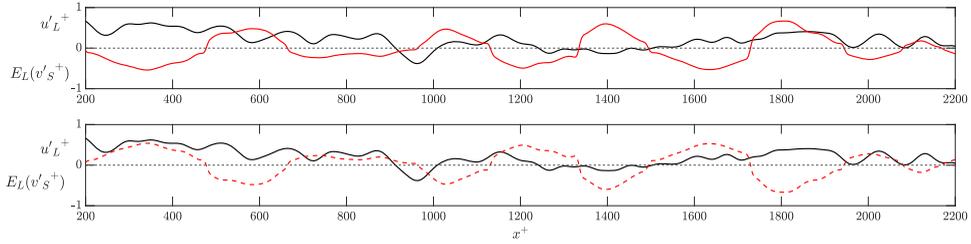

Figure D.2: The top plot shows large-scale streamwise velocity fluctuations (——) at $y^+ \approx$ +100 and the long-wavelength envelope of small-scale wall-normal velocity fluctuations (——) at $y^+ = 0$ of case $HP1$; the bottom figure shows the same, only with the envelope (– –) now phase-shifted by $\pi$.

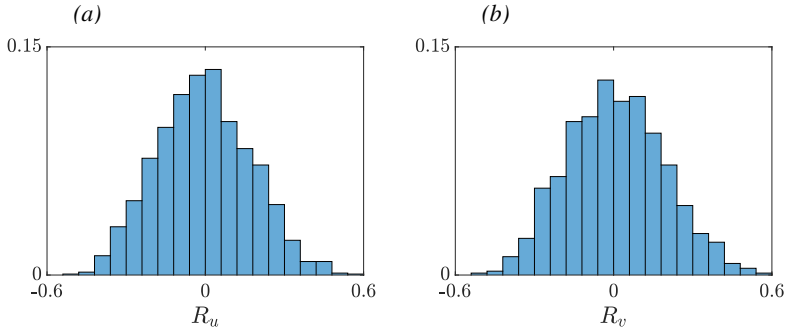

Figure D.3: Probability density histograms of AM correlation: (a) between $u'L^+$ and $E_L(u'_s{}^+)$, (b) between $u'L^+$ and $E_L(u'_s{}^+)$. The $u'$ signal was taken at $y^+ \approx 100$ while the $v'$ signals was taken at $y = 0$.

## REFERENCES

ABDERRAHAMAN-ELENA, N., FAIRHALL, C.T. & GARCÍA-MAYORAL, R. 2019 Modulation of near-wall turbulence in the transitionally rough regime. *Journal of Fluid Mechanics* **865**, 1042–1071.

ANISZEWSKI, W., ARRUFAT, T, CRIALESI-ESPOSITO, M, DABIRI, S, FUSTER, D, LING, Y, LU, J, MALAN, L, PAL, S, SCARDOVELLI, R, TRYGGVASON, G, YECKO, P & ZALESKI, Stéphane 2019 Parallel, robust, interface simulator (paris). Working paper or preprint.

BAARS, WJ, TALLURU, KM, HUTCHINS, Nicholas & MARUSIC, Ivan 2015 Wavelet analysis of wall turbulence to study large-scale modulation of small scales. *Experiments in Fluids* **56** (10), 1–15.

BREUGEM, W. P. & BOERSMA, B. J. 2005 Direct numerical simulations of turbulent flow over a permeable wall using a direct and a continuum approach. *Physics of Fluids* **17** (2), 025–103.

BREUGEM, W. P., BOERSMA, B. J. & UITTENBOGAARD, R. E. 2006 The influence of wall permeability on turbulent channel flow. *Journal of Fluid Mechanics* **562**, 35–72.

CHEN, ZISHEN & GARCÍA-MAYORAL, RICARDO 2023 Examination of outer-layer similarity in highly obstructed wall turbulence: canopy flows. *arXiv preprint arXiv:2305.16764* .

CHUNG, D., HUTCHINS, N., SCHULTZ, M. P. & FLACK, K. A. 2021 Predicting the drag of rough surfaces. *Annual Review of Fluid Mechanics* **53** (1), 439–471.

CLAUSER, FRANCIS H 1954 Turbulent boundary layers in adverse pressure gradients. *Journal of the Aeronautical Sciences* **21** (2), 91–108.

COSTA, PEDRO 2018 A fft-based finite-difference solver for massively-parallel direct numerical simulations of turbulent flows. *Computers & Mathematics with Applications* **76** (8), 1853 – 1862.

EFSTATHIOU, CHRISTOPH & LUHAR, MITUL 2018 Mean turbulence statistics in boundary layers over high-porosity foams. *Journal of Fluid Mechanics* **841**, 351–379.

FINNIGAN, J. 2000 Turbulence in plant canopies. *Annual Review of Fluid Mechanics* **32** (1), 519–571.

FINNIGAN, JOHN J., SHAW, ROGER H. & PATTON, EDWARD G. 2009 Turbulence structure above a vegetation canopy. *Journal of Fluid Mechanics* **637**, 387–424.

GARCÍA-MAYORAL, R., GÓMEZ-DE SEGURA, G. & FAIRHALL, C.T. 2019 The control of near-wall turbulence through surface texturing. *Fluid Dynamics Research* **51** (1), 011410.

HABIBI KHORASANI, SEYED MORTEZA, LĀCIS, UĢIS, PASCHE, SIMON, ROSTI, MARCO EDOARDO & BAGHERI, SHERVIN 2022 Near-wall turbulence alteration with the transpiration-resistance model. *Journal of Fluid Mechanics* **942**, A45.

HAMA, FRANCIS R 1954 Boundary layer characteristics for smooth and rough surfaces. *Trans. Soc. Nav. Arch. Marine Engrs.* **62**, 333–358.

HASSAN, Y. A. & DOMINGUEZ-ONTIVEROS, E.E. 2008 Flow visualization in a pebble bed reactor experiment using piv and refractive index matching techniques. *Nuclear Engineering and Design* **238** (11), 3080–3085, hTR-2006: 3rd International Topical Meeting on High Temperature Reactor Technology.

IBRAHIM, J.I., GÓMEZ-DE-SEGURA, G., CHUNG, D. & GARCÍA-MAYORAL, R. 2021 The smooth-wall-like behaviour of turbulence over drag-altering surfaces: a unifying virtual-origin framework. *Journal of Fluid Mechanics* **915**, A56.

JIMÉNEZ, J. 1994 On the structure and control of near wall turbulence. *Physics of Fluids* **6** (2), 944–953.

JIMÉNEZ, J., UHLMANN, M., PINELLI, A. & KAWAHARA, G. 2001 Turbulent shear flow over active and passive porous surfaces. *Journal of Fluid Mechanics* **442**, 89–117.

KAZEMIFAR, F., BLOIS, G., AYBAR, M., PEREZ CALLEJA, P., NERENBERG, R., SINHA, S., HARDY, R. J., BEST, J., SAMBROOK S., GREGORY H. & CHRISTENSEN, K. T. 2021 The effect of biofilms on turbulent flow over permeable beds. *Water Resources Research* **57** (2), e2019WR026032, e2019WR026032 2019WR026032.

KIM, J. & MOIN, P. 1985 Application of a fractional-step method to incompressible navier-stokes equations. *Journal of Computational Physics* **59** (2), 308–323.

KIM, T., BLOIS, G., BEST, J. L. & CHRISTENSEN, K. T. 2020 Experimental evidence of amplitude modulation in permeable-wall turbulence. *Journal of Fluid Mechanics* **887**, A3.

KURUNERU, S. T. W., VAFAI, K., SAURET, E. & GU, Y. 2020 Application of porous metal foam heat exchangers and the implications of particulate fouling for energy-intensive industries. *Chemical Engineering Science* **228**, 115968.

KUWATA, Y. & SUGA, K. 2016 Lattice boltzmann direct numerical simulation of interface turbulence over porous and rough walls. *International Journal of Heat and Fluid Flow* **61**, 145–157, sI TSFP9 special issue.

KUWATA, Y. & SUGA, K. 2017 Direct numerical simulation of turbulence over anisotropic porous media. *Journal of Fluid Mechanics* **831**, 41–71.

LEE, MYOUNGKYU & MOSER, ROBERT D. 2015 Direct numerical simulation of turbulent channel flow up to $Re_\tau \approx 5200$. *Journal of Fluid Mechanics* **774**, 395–415.

LUCHINI, P 1996 Reducing the turbulent skin friction. In *Computational methods in applied sciences' 96 (Paris, 9-13 September 1996)*, pp. 465–470. John Wiley & Sons Ltd.

LUCHINI, P. 2015 The relevance of longitudinal and transverse protrusion heights for drag reduction by a superhydrophobic surface. In *Proc. European Drag Reduction and Flow Control Meeting—EDRFMC 2015; March 23–26*, pp. 81–82.

LUCHINI, P., MANZO, F. & POZZI, A. 1991 Resistance of a grooved surface to parallel flow and cross-flow. *Journal of Fluid Mechanics* **228**, 87–109.

MACDONALD, M., CHAN, L., CHUNG, D., HUTCHINS, N. & OOI, A. 2016 Turbulent flow over transitionally rough surfaces with varying roughness densities. *Journal of Fluid Mechanics* **804**, 130–161.

MANES, C., POGGI, D. & RIDOLFI, L. 2011 Turbulent boundary layers over permeable walls: scaling and near-wall structure. *Journal of Fluid Mechanics* **687**, 141–170.

MANES, C., POKRAJAC, D., MCEWAN, I. & NIKORA, V. 2009 Turbulence structure of open channel flows over permeable and impermeable beds: A comparative study. *Physics of Fluids* **21** (12), 125109.

MATHIS, R., HUTCHINS, N. & MARUSIC, I. 2009 Large-scale amplitude modulation of the small-scale structures in turbulent boundary layers. *Journal of Fluid Mechanics* **628**, 311–337.

ORLANDI, P. & LEONARDI, S. 2006 Dns of turbulent channel flows with two- and three-dimensional roughness. *Journal of Turbulence* **7**, N73.

PARAVENTO, F., POURQUIE, M. J. & BOERSMA, B. J. 2008 An immersed boundary method for complex flow and heat transfer. *Flow, Turbulence and Combustion* **80** (2), 187–206.

REYNOLDS, W. C. & HUSSAIN, A. K. M. F. 1972 The mechanics of an organized wave in turbulent shear flow. part 3. theoretical models and comparisons with experiments. *Journal of Fluid Mechanics* **54** (2), 263–288.

ROSTI, MARCO E., BRANDT, LUCA & PINELLI, ALFREDO 2018 Turbulent channel flow over an anisotropic porous wall – drag increase and reduction. *Journal of Fluid Mechanics* **842**, 381–394.

GÓMEZ-DE SEGURA, G. & GARCÍA-MAYORAL, R. 2019 Turbulent drag reduction by anisotropic permeable substrates – analysis and direct numerical simulations. *Journal of Fluid Mechanics* **875**, 124–172.

GÓMEZ-DE SEGURA, G., SHARMA, A. & GARCÍA-MAYORAL, R. 2018 Turbulent drag reduction using anisotropic permeable substrates. *Flow, Turbulence and Combustion* **100** (4), 995–1014.

SHAHZAD, HARIS, HICKEL, STEFAN & MODESTI, DAVIDE 2023 Turbulence and added drag over acoustic liners. *Journal of Fluid Mechanics* **965**, A10.

SHARMA, A. & GARCÍA-MAYORAL, R. 2020 Turbulent flows over dense filament canopies. *Journal of Fluid Mechanics* **888**, A2.

SHARMA, AKSHATH, GOMEZ-DE SEGURA, GARAZI & GARCIA-MAYORAL, RICARDO 2017 Linear stability analysis of turbulent flows over dense filament canopies. In *Tenth International Symposium on Turbulence and Shear Flow Phenomena*. Begel House Inc.

SHEN, GUANGCHEN, YUAN, JUNLIN & PHANIKUMAR, MANTHA S. 2020 Direct numerical simulations of turbulence and hyporheic mixing near sediment–water interfaces. *Journal of Fluid Mechanics* **892**, A20.

SPALART, P.R. & MCLEAN, J.D. 2011 Drag reduction: enticing turbulence, and then an industry. *Philosophical Transactions of the Royal Society A: Mathematical, Physical and Engineering Sciences* **369** (1940), 1556–1569.

SUNDIN, JOHAN, ZALESKI, STÉPHANE & BAGHERI, SHERVIN 2021 Roughness on liquid-infused surfaces induced by capillary waves. *Journal of Fluid Mechanics* **915**, R6.

TOWNE, AARON, SCHMIDT, OLIVER T. & COLONIUS, TIM 2018 Spectral proper orthogonal decomposition and its relationship to dynamic mode decomposition and resolvent analysis. *Journal of Fluid Mechanics* **847**, 821–867.

WANG, WENKANG, LOZANO-DURÁN, ADRIÁN, HELMIG, RAINER & CHU, XU 2022 Spatial and spectral characteristics of information flux between turbulent boundary layers and porous media. *Journal of Fluid Mechanics* **949**, A16.

WHITE, BRIAN L. & NEPF, HEIDI M. 2007 Shear instability and coherent structures in shallow flow adjacent to a porous layer. *Journal of Fluid Mechanics* **593**, 1–32.

ZAGNI, A. F. E. & SMITH, K. V. H. 1976 Channel flow over permeable beds of graded spheres. *Journal of the Hydraulics Division* **102** (2), 207–222.